\RequirePackage{lineno}
\documentclass[amsmath, amssymb,10pt,aps,prb,twocolumn,notitlepage,showpacs,superscriptaddress]{revtex4-1}
\usepackage{graphicx} 
\usepackage[bookmarks=false]{hyperref}
\usepackage{amsmath}
\usepackage{amssymb}
\usepackage{bm}
\usepackage{amsthm}
\usepackage{amsfonts}
\usepackage{latexsym}
\usepackage{wrapfig}
\usepackage[usenames,dvipsnames]{color}
\usepackage[protrusion=true,expansion=true,final]{microtype}
\usepackage{ifthen}
\usepackage{tikz}\usetikzlibrary{shapes,arrows,positioning}
\usepackage{color,soul}
\usepackage{lineno}
\usepackage{csquotes}
\usepackage{booktabs}
\usepackage[normalem]{ulem}
\usepackage{chemformula}
\useunder{\uline}{\ul}{}
\usepackage{textcomp}
\usepackage{accents}

\usepackage{amsmath,scalerel}

\DeclareMathOperator*{\Bigcdot}{\scalerel*{\cdot}{\bigodot}}

\usepackage{amsmath,amssymb,mathtools}
\usepackage{bm}
\usepackage{siunitx}
\usepackage{graphicx}
\usepackage{booktabs}
\usepackage{physics}
\usepackage[most]{tcolorbox}
\usepackage{microtype}
\usepackage{cleveref}

\newcommand{\avg}[1]{\left\langle #1 \right\rangle}

\begin{document}

\title{Field-unmasked quantum geometry in a symmetry-forbidden photocurrent}
\author{Bumseop Kim}
\thanks{These authors contributed equally.}
\affiliation{Department of Chemistry, University of Pennsylvania, Philadelphia, Pennsylvania 19104, USA}

\author{Aaron M. Burger}
\thanks{These authors contributed equally.}
\affiliation{Department of Electrical and Computer Engineering, Drexel University, Philadelphia, Pennsylvania 19104, USA}

\author{Zhenbang Dai}
\affiliation{Oden Institute for Computational Engineering and Sciences, The University of Texas at Austin, Austin, Texas 78712, USA}
\affiliation{Department of Physics, The University of Texas at Austin, Austin, Texas 78712, USA}

\author{Sayed Ali Akbar Ghorashi}
\affiliation{Department of Chemistry, University of Pennsylvania, Philadelphia, Pennsylvania 19104, USA}

\author{Adam Abirou}
\affiliation{Department of Physics, Drexel University, Philadelphia, Pennsylvania 19104, USA}

\author{Md Al Helal}
\affiliation{Department of Materials Science and Engineering, Drexel University, Philadelphia, Pennsylvania 19104, USA}

\author{Vladimir M. Fridkin}
\affiliation{Department of Physics, Drexel University, Philadelphia, Pennsylvania 19104, USA}
\affiliation{Shubnikov Institute for Crystallography, Russian Academy of Sciences, Leninsky Prospect 59, Moscow 117333, Russia}

\author{Jonathan E. Spanier}
\affiliation{Department of Electrical and Computer Engineering, Drexel University, Philadelphia, Pennsylvania 19104, USA}
\affiliation{Department of Physics, Drexel University, Philadelphia, Pennsylvania 19104, USA}
\affiliation{Department of Materials Science and Engineering, Drexel University, Philadelphia, Pennsylvania 19104, USA}

\author{Andrew M. Rappe}
\email{rappe@sas.upenn.edu}
\affiliation{Department of Chemistry, University of Pennsylvania, Philadelphia, Pennsylvania 19104, USA}

\date{\today}

\begin{abstract}
Frequency- and polarization-resolved photocurrents provide a sensitive probe of hidden symmetry and band geometry in quantum materials~\cite{zhang2019Nature, akamatsu2021Science, young2012first, zhang2019NatCommun, morimoto2016SciAdv, ahn2022NatPhys, Ma2021, Kang2025, krishna2025NatMater}. Here we study a chiral cubic sillenite, \ch{Bi12SiO20}, whose global crystal symmetry forbids a longitudinal odd-in-$B$ magneto-photocurrent in the Voigt geometry~\cite{fridkin1985magnetophotovoltaic}. Nevertheless, we observe a pronounced longitudinal response across the visible range that is predominantly linear in magnetic field, persists below the band gap, and exhibits strong helicity selectivity, with the circular channel exceeding the linear one and reversing sign upon switching light helicity. We resolve this apparent contradiction by identifying defect-enabled, field-selected spin ordering as the mechanism that lowers the effective magnetic symmetry without altering the global crystal structure. First-principles calculations show that oxygen vacancies generate in-gap bound states and localized magnetic moments on neighboring Bi-O units, stabilized by strong spin-orbit coupling. Although symmetry-related vacancy configurations remain energetically degenerate and preserve the macroscopic $T$ symmetry at zero field, an applied magnetic field selects a time-reversal-broken sector of the defect ensemble and reduces the effective magnetic symmetry to the subgroup that leaves $\hat{B}$ invariant, thereby lifting the longitudinal selection rule. Importantly, this field-selected symmetry reduction does more than activate a nominally forbidden photocurrent: it unmasks latent quantum-geometric responses encoded in the electronic structure. Momentum-resolved calculations show that the dominant circular and linear magneto-photocurrent channels spatially correlate with Berry-curvature-rich and quantum-metric-rich regions of the Brillouin zone, respectively. Our results establish field-selected defect symmetry lowering as a route to revealing hidden quantum geometry and activating forbidden nonlinear photocurrents in chiral quantum materials.
\end{abstract}

\maketitle

\section{Introduction}

Photocurrents provide a powerful probe of hidden symmetries and related electronic structure in solids~\cite{grinberg2013Nature, mak2010PRL, graham2013NatPhys, akamatsu2021Science, zhang2019Nature}. The direction of the current is dictated by the underlying spatial symmetries, whereas its magnitude is controlled by the band topology and quantum geometry, including the Berry curvature and quantum metric of Bloch states. It is therefore highly sensitive to the interplay among crystalline symmetry, band topology, and external perturbations such as electromagnetic fields~\cite{ma2017NatPhys, ma2022Nature, xu2018NatPhys}. In particular, the polarization dependence of photocurrents gives direct access to selection rules, enabling the separation of microscopic contributions from intrinsic band-geometry effects and extrinsic mechanisms~\cite{mciver2012NatNanotechnol, yang2018AdvMater, matsubara2022NatCommun}. Recent works have leveraged these features to probe Berry curvature and topological band structure via nonlinear optical responses, including the bulk photovoltaic effect (BPVE) and the circular photogalvanic effect~\cite{ji2019NatureMater, morimoto2016SciAdv, ahn2020PRX, zhang2019NatCommun}. These developments highlight photocurrent spectroscopy as both an experimental diagnostic of purely electronic states that are difficult to access with conventional spectroscopies~\cite{ma2023NatRevPhys, sirica2022NatMater, pettine2023Light} and a means to reveal hidden material properties~\cite{Wang2024_NatComm_hiddenSpinCPC,Cheon2023_npj2D_BPVE_PtSe2}.

Among nonlinear photocurrent phenomena, the BPVE, a second-order current under uniform illumination, is particularly sensitive to crystal symmetry and attractive for next-generation optoelectronics~\cite{koch1976Ferro, Young:2012, spanier2016NatPhoton, aftab2022AdvOptMater}. In an external magnetic field, additional magneto-optical structure emerges and enables magnetic control over light–matter coupling~\cite{Burger:2019, Burger:2020, zhang2019NatCommun, dai2023PRB, pi2023magnetic}. Generically, the second-order current density can be written in powers of the electric field \(\mathbf{E}\) and the magnetic field \(\mathbf{B}\) as,
\begin{equation}
\begin{aligned}
J_i(\vb E,\vb B)
&=\sum_{j\ge 0}
\chi^{(2,j)}_{\,i;\,(\alpha_1\alpha_2);(\beta_1\cdots\beta_j)}
E_{\alpha_1}E_{\alpha_2}
\left(\prod_{l=1}^{j} B_{\beta_l}\right), \\
&\qquad i,\alpha_r,\beta_l\in\{x,y,z\}.
\end{aligned}
\label{MPVE}
\end{equation}
which makes symmetry constraints on the tensor coefficients explicit~\cite{fridkin1985magnetophotovoltaic}.

In high-symmetry crystals such as chiral cubic point group $T$ [Fig.~\ref{fig1}(a)], these constraints can tightly restrict responses. In particular, a longitudinal odd-in-$B$ magneto-photocurrent in the Voigt geometry, \(\mathbf{J}\parallel\mathbf{k}\perp\mathbf{B}\) where \(\mathbf{k}\) is the light-propagation direction, is forbidden by $T$-symmetry selection rules [Fig.~\ref{fig1}(b)]~\cite{fridkin1985magnetophotovoltaic}. Thus, the response under odd powers of the magnetic field in Eq.~\ref{MPVE} should vanish. Bismuth-containing sillenite crystals with point group $T$, such as \ch{Bi_{12}SiO_{20}} (BSO), have long been studied for their strong photorefractive, electro-optic, and nonlinear optical responses~\cite{Tassev:2000, Mihailovic:2008}, but their application has been limited by their symmetry constraints. Contrary to this expectation, we observe significant longitudinal magneto-photovoltaic effect (MPVE) across the entire visible range (325-647 nm), including sub-band-gap excitation and predominantly linear order in \(B\). These observations point to an additional ingredient that modifies the effective symmetry relevant to the magneto-optical response beyond that of the ideal crystal.

In this work, we show that oxygen-vacancy defects enable a field-selected spin-ordering mechanism that lowers the effective magnetic symmetry while preserving the global crystallographic point group $T$~\cite{Burkov:2001, ahmad2009, Reyher:1992, Ivchenko:2017}. Structural defects, which are ubiquitous in chiral high-symmetry BSO, locally reduce rotational symmetry and create in-gap states and unpaired spin moments, while their symmetry-related configurations remain statistically uniform at $B=0$ and therefore preserve macroscopic $T$ symmetry. Experimentally, we observe pronounced sub-band-gap longitudinal photocurrents with clear polarization harmonics and strong helicity selectivity. We find that vacancies generate localized magnetic moments on neighboring Bi-O units stabilized by strong Bi spin-orbit coupling, and that an applied magnetic field selects among spin-orbit-coupled magnetocrystalline-anisotropy minima, introducing a distinct axial direction and reducing the effective magnetic symmetry to the subgroup that leaves $\hat{B}$ invariant, $T \rightarrow C_{2\hat{B}}$. This field-selected symmetry lowering lifts the longitudinal selection rule and activates the odd-in-$B$ magneto-photocurrent channel observed experimentally. More importantly, the activated response provides access to latent quantum geometry that is hidden by symmetry in the pristine crystal. By directly connecting the dominant circular and linear magneto-photocurrent channels to Berry-curvature-rich and quantum-metric-rich regions of momentum space, respectively, we show that defect- and field-selected symmetry lowering does not merely unlock a forbidden response, but also unmasks the underlying band geometry. Together, these results establish a unified mechanism for field-unmasking hidden quantum geometry and activating symmetry-forbidden nonlinear photocurrents in chiral quantum materials.

\section{Results}


\subsection*{\textbf{Forbidden-to-allowed longitudinal magneto-photocurrent in chiral BSO.}}

\begin{figure*}
    \centering
    \includegraphics[width=0.7\textwidth]{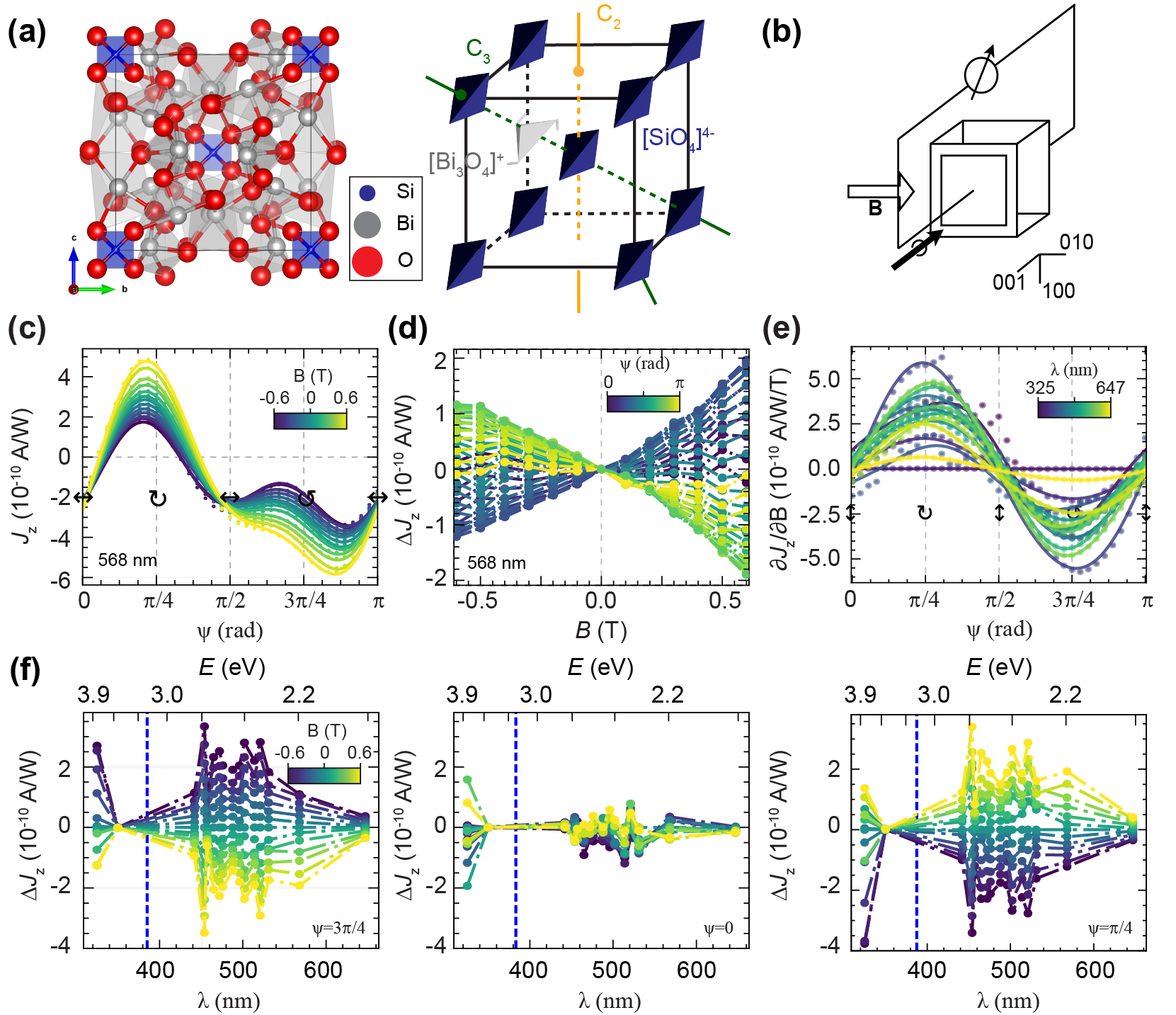}
    \caption{
    (a) Crystal structure of Bi$_{12}$SiO$_{20}$: atomic model (Si blue, Bi grey, and O red) and a simplified tetrahedral diagram highlighting its chiral cubic symmetry $T$ with two-fold and three-fold symmetry (space group $I23$).
    (b) Measurement geometry in the Voigt configuration for longitudinal photocurrent under magnetic field: Photocurrent $\mathbf{J} \parallel \hat{z}$, magnetic field $\mathbf{B} \parallel \hat{y}$, and light propagates along $\mathbf{k} \parallel \hat{z}$.
    (c) Magneto-optical photocurrent $J_{z}$ versus quarter-wave-plate angle $\psi$ at $\lambda=568$~nm; color encodes magnetic field $B$ (T; $+\hat{y}$ positive; $-0.6 \rightarrow 0.6$ in 0.1 T steps).
    (d) Magnetic field dependence of $J_z$ ($\Delta J_z$) at $\lambda=568$~nm for quarter-wave plate angles $\psi\in[0,\pi]$.
    (e) Magnetic sensitivity $\partial J_z/\partial B$ versus light-polarization state for all measured wavelengths.
    (f) Spectra of the magneto-photocurrent change $\Delta J_{z}(B)\equiv J_{z}(B)-J_{z}(B=0)$ for left circularly (left), linearly (middle), and right circularly (right) polarized light. The blue dashed line marks the pristine BSO band gap ($\approx 387$~nm).}
    \label{fig1}
\end{figure*}

BSO is a chiral, non-centrosymmetric crystal in the cubic point group $T$ and space group $I23$, comprising a central [SiO$_4$]$^{4-}$ tetrahedron surrounded by a cage-like [Bi$_3$O$_4$]$^{+}$ framework. The structure admits two-fold and three-fold rotations, which invert transverse Cartesian components and cyclically permute the axes, respectively as shown in Fig.~\ref{fig1}(a). 
The second-order magneto-optical current can be written as,
\begin{equation}
J_a = S_{abcd}\, E_b E_c^*\,B_d  + i\, Q_{abc}\, (\mathbf{E} \times \mathbf{E}^*)_b\, B_c,
\label{secondMPVE}
\end{equation}
where $S_{abcd}$ and $Q_{abc}$ are the fourth and third-rank tensors characterizing the linear and circular MPVE responses, respectively~\cite{sturman1992photovoltaic, fridkin1985magnetophotovoltaic, Hornung:2021}. They are allowed only when the light-propagation vector $\mathbf{k}$, the photocurrent direction $\mathbf{J}$, and the magnetic field direction $\mathbf{B}$ are orthogonal to each other in point group $T$; any longitudinal component reverses sign under the crystal’s two-fold rotations while $\mathbf{k}$ remains invariant—a selection rule confirmed experimentally~\cite{petrov1979photogalvanic}.

We probe magneto-photocurrent in the Voigt geometry with $\mathbf{B} \parallel \hat{\mathbf{y}}$ and $\mathbf{k} \parallel \hat{\mathbf{z}}$ [Fig.~\ref{fig1}(b)]. Symmetry analysis (full tensor derivation is provided in the Supplementary Note ~\ref{SI.NoteS1.A} and ~\ref{SI.NoteS1.B}) shows that odd-in-$B$ magneto-optical terms (i.e., odd-$q$ terms in Eq.~\ref{MPVE}) are allowed only for Hall configurations ($\mathbf{J} \perp \mathbf{k}$), whereas even-in-$B$ terms (even-$q$ terms in Eq.~\ref{MPVE}) are only allowed for longitudinal components ($\mathbf{J} \parallel \mathbf{k}$). In our experiment, contrary to the symmetry analysis, we observe a longitudinal $B$-linear order magneto-photocurrent ($\mathbf{J}\parallel\mathbf{k}$) in BSO across the visible range (325-647 nm) while rotating a quarter-wave plate (QWP) by an angle $\psi$ [Fig.~\ref{fig1}(c) and Fig.~\ref{figSI3}]. A rotating QWP transforms the light from linear $xy$-polarized ($\psi=0$) to right- and left-circular polarization (RCP, $\psi=\pi/4$; LCP, $\psi=3\pi/4$)~\cite{Grachev:1982}, thereby decomposing the response into helicity-odd and linear-polarization harmonics. The measured currents $J_z(\psi)$ are well fitted by~\cite{Ivchenko:2017, ji2019NatureMater}

\begin{equation}
J_z(\psi) = J_L \sin(4\psi+\psi_0) -J_C\sin(2\psi) +J_{\mathrm{0}}
\label{J_fitting}
\end{equation}
where the magnetic-field dependence of $J_L$ and $J_C$ encode the linear and circular photocurrent tensor component, respectively. $J_{\mathrm{0}}$ denotes a polarization-independent current, which typically occurs near the electrodes and vanishes toward the midpoint.

Four salient observations emerge. (i) The longitudinal differential signal $\Delta J_z \equiv J_z(B)-J_z(0)$ exhibits a clear linear-in-$B$ dependence in the Voigt geometry [Fig.~\ref{fig1}(d) and Fig.~\ref{figSI4}], which we argued is forbidden by $T$ symmetry in the longitudinal channel. The measured current change reverses sign when the $B$ field is reversed, indicating the response is an odd function of the $B$ field. We quantify magnetic sensitivity by extracting the slope $\partial J/\partial B$ for each $\psi$ from linear fits to the traces in Fig.~\ref{fig1}(d). The experimental sensitivity peaks for RCP $(\psi=\pi/4)$ and LCP $(\psi=3\pi/4)$ and reverses sign between the two across the visible range [Fig.~\ref{fig1}(e) and Fig.~\ref{figSI5}]. (ii) Across the visible spectrum, the magnetophotovoltaic response under circularly-polarized light is approximately four times larger than the one with linear polarization [Fig.~\ref{fig1}(f)]. (iii) Reversing helicity (RCP $\leftrightarrow$ LCP) flips the sign of $J_z$ at fixed $B$ [Fig.~\ref{fig1}(e) and ~\ref{fig1}(f)]. (iv) Strong signals persist well below the experimental band gap of $\approx 3.2$\,eV (blue dashed line in Fig.~\ref{fig1}(f))~\cite{hou1973transport, isik2021defect}. Taken together, these observations imply that generating a second-order photocurrent at sub-band-gap photon energies requires a mechanism beyond the conventional symmetry arguments.


\subsection*{\textbf{Defect-enabled magneto-electric response in Bi$_{12}$SiO$_{20}$.}}

\begin{figure*}
    \centering
    \includegraphics[width=0.4\textwidth]{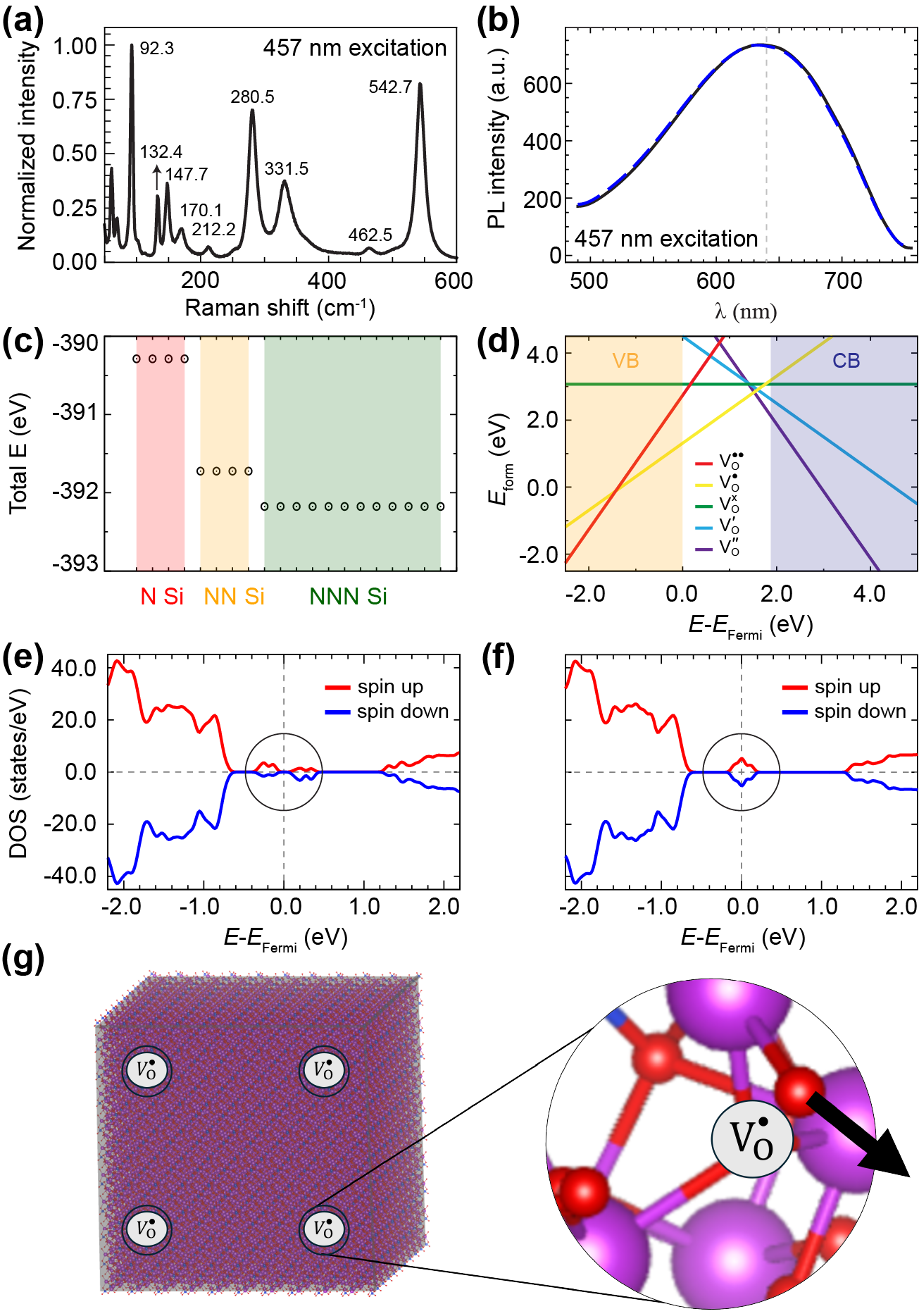}
\caption{
(a) Room-temperature Raman spectrum of \ch{Bi12SiO20} collected via a back-scattering configuration under 457 nm excitation with a parallel polarization geometry. Raman spectra exhibit peaks at 61.0, 69.6, 92.3, 132.4, 147.7, 170.1, 212.2, 280.5, 331.5, 462.5, and 542.7 cm$^{-1}$.
(b) Photoluminescence spectrum of \ch{Bi12SiO20} with an excitation wavelength of 457 nm depending on helicity of light (RCP, black-solid; LCP blue-dashed lines).
(c) Total energies of symmetry-related neutral oxygen-vacancy configurations $\left(\mathrm{V}_{\mathrm O}^{\times}\right)$, grouped by the distance to the nearest Si site (N/NN/NNN).
(d) Defect formation energy of \ch{Bi12SiO20} with an oxygen vacancy as a function of the Fermi level; the slope of each segment equals the charge state $q$. Shaded regions indicate the valence-band (VB) and conduction-band (CB) energy ranges.
(e,f) Spin-resolved density of states (DOS) for \ch{Bi12SiO20} containing $\mathrm{V}_{\mathrm O}^{\Bigcdot}$: (e) a magnetic solution with total magnetization $M \approx 1\,\mu_{\mathrm B}$ per supercell and (f) a (nearly) nonmagnetic solution with $M \approx 0$, highlighting a defect-related feature near $E_{\mathrm F}$ (circled).
(g) Supercell view and local atomic environment around $\mathrm{V}_{\mathrm O}^{\Bigcdot}$, illustrating vacancy-induced spin polarization on neighboring Bi atoms (arrow indicates the spin/magnetization direction).
}
    \label{fig2}
\end{figure*}

Because we measure a second-order photocurrent at photon energies below the BSO band gap, the response necessarily involves in-gap states. Since BSO is also known to host a variety of defect-related optical phenomena~\cite{hou1973transport, isik2021defect, Lima:2011, lima2013first, ahmad2009, Burkov:2001, lauer1971thermally}, the most parsimonious interpretation is that defect-induced properties enable the longitudinal, odd-in-$B$, helicity-selective magneto-photocurrent channel observed here. To further examine this picture, we first measured room-temperature Raman spectra. As shown in Fig.~\ref{fig2}(a), the observed Raman peaks match the characteristic phonon modes reported for sillenite BSO, supporting the phase purity of the measured crystals~\cite{wojdowski1985vibrational,lazarevic2013raman,isik2019temperature}. We then performed photoluminescence (PL) measurements, shown in Fig.~\ref{fig2}(b). Under 457 nm circularly polarized excitation, BSO exhibits a PL band centered near 640 nm, in good agreement with earlier reports~\cite{cremades1996cathodoluminescence}. This PL response is consistent with defect-mediated recombination through optically active in-gap states. Taken together, these spectroscopic results support the defect-based interpretation of the sub-band-gap photocurrent. Among the possible defect species, oxygen vacancies are widely considered to be among the most abundant under oxygen-poor conditions~\cite{Burkov:2001}.

Our calculations show that the total energy of O vacancies under magnetic field across symmetry-related sites are degenerate for sites at the same Si proximity (N/NN/NNN Si) [Fig.~\ref{fig2}(c)], implying a statistically uniform occupation that preserves the global $T$ symmetry. Hence, each individual vacancy lowers symmetry locally, while the ensemble retains the crystalline point group, consistent with our group-theoretical analysis (see Supplementary Note ~\ref{SI.NoteS1.C}). Therefore, since the defects themselves do not break the spatial symmetry, the observed symmetry-forbidden magneto-photocurrent must arise from how these defects modify the system’s response beyond what is captured by the ideal crystal symmetry.

One well-known defect-related phenomenon is defect-induced magnetism, and even $p$-orbital-based graphene shows sizable moments upon defect creation~\cite{nair2012spin,yazyev2010emergence}. As shown in Fig.~\ref{fig2}(d), removing O leaves an unpaired electron depending on charge state. This magnetic moment adopts orientations stabilized by the strong spin-orbit coupling (SOC) of Bi~\cite{brand2025APR}. Details of the charged defect calculation are described in the Methods section. Consistent with this picture, previous electron paramagnetic resonance (EPR) measurements report an effective spin-$1/2$ center upon vacancy formation~\cite{Sekhar2012BSOThesis}. We summarized defect formation energy calculations in the Methods section. In our DFT calculations, allowing noncollinear spin polarization stabilizes the defect state; compared to a spin-constrained solution, the spinor wavefunction solution opens a spin-dependent gap, as shown in Fig.~\ref{fig2}(e) and (f), and lowers the total energy by approximately 70~meV.

To determine in detail how a charged oxygen vacancy sets a preferred spin orientation in the absence of an external field, we map the defect magnetocrystalline anisotropy energy (MAE) induced by Bi spin-orbit coupling. For each symmetry-related vacancy configuration $p$, we constrain the magnetization direction $\hat{\vb m}$ and compute the total energy $E^{(p)}(\hat{\vb m})$ over a quasi-uniform set of orientations on $\mathbb{S}^2$ (Fibonacci sampling, including antipodal partners to enforce $E(\hat{\vb m})=E(-\hat{\vb m})$ at $B=0$). We then extract the easy axis $\hat{\vb m}^{(p)}_{\mathrm{easy}}$ using a robust leading-anisotropy quadratic fit $E^{(p)}(\hat{\vb m}) = C^{(p)} + \hat{\vb m}^{\mathsf T} M^{(p)} \hat{\vb m}$, where the minimum-eigenvalue eigenvector of the symmetric matrix $M^{(p)}$ defines $\hat{\vb m}^{(p)}_{\mathrm{easy}}$. Across symmetry-related defect sites, the resulting easy axes transform covariantly under crystallographic rotations, $\hat{\vb m}^{(q)}_{\mathrm{easy}} \simeq R\,\hat{\vb m}^{(p)}_{\mathrm{easy}}$, consistent with the underlying $T$ symmetry of the ionic lattice. Detailed information for the spin easy axis is summarized in Supplementary Note ~\ref{SI.NoteS2}.

At $B=0$, the nearly degenerate formation energies of symmetry-related vacancy sites imply a statistically uniform configurational ensemble, so the net magnetization cancels and the macroscopic point group remains $T$ (and space group $I23$). Applying a magnetic field selects among the SOC-coupled MAE minima by favoring the branch with the local moment aligned with $\hat{\vb B}$ (in our Voigt geometry, $\vb B\parallel \hat{\vb y}$), thereby introducing a distinguished axial direction in the electronic sector even without any ionic symmetry breaking. Consequently, the effective symmetry is reduced to the subgroup of $T$ that leaves $\hat{\vb B}$ invariant, i.e., $T \rightarrow \{E, C_{2\hat{\vb B}}\}\equiv C_2$. This field-selected magnetic point-group reduction removes the cancellations enforced by the orthogonal $C_2$ axes in $T$ and therefore opens a longitudinal photocurrent channel that is odd in $\vb B$ and linear in $B$ at leading order, in agreement with the observed Voigt-geometry response.

This mechanism resolves the full set of experimental anomalies within a single picture. Oxygen vacancies introduce in-gap states, thereby enabling sub-band-gap photoexcitation, while simultaneously generating localized spin moments stabilized by strong SOC. At zero field, symmetry-related vacancy configurations are degenerate and statistically preserve the macroscopic $T$ symmetry, so no net symmetry-forbidden longitudinal odd-in-$B$ response is selected. Under an applied magnetic field, however, the defect ensemble is projected onto the spin-orbit-coupled branch favored by $\hat{\mathbf B}$, which lowers the effective magnetic point group to the subgroup compatible with the field direction. This field-selected symmetry reduction removes the cancellation enforced in the pristine crystal and activates the longitudinal magneto-photocurrent channel observed experimentally.


\subsection*{\textbf{Defect-activated longitudinal magneto-photocurrent in BSO.}}

\begin{figure*}
    \centering
    \includegraphics[width=0.9\textwidth]{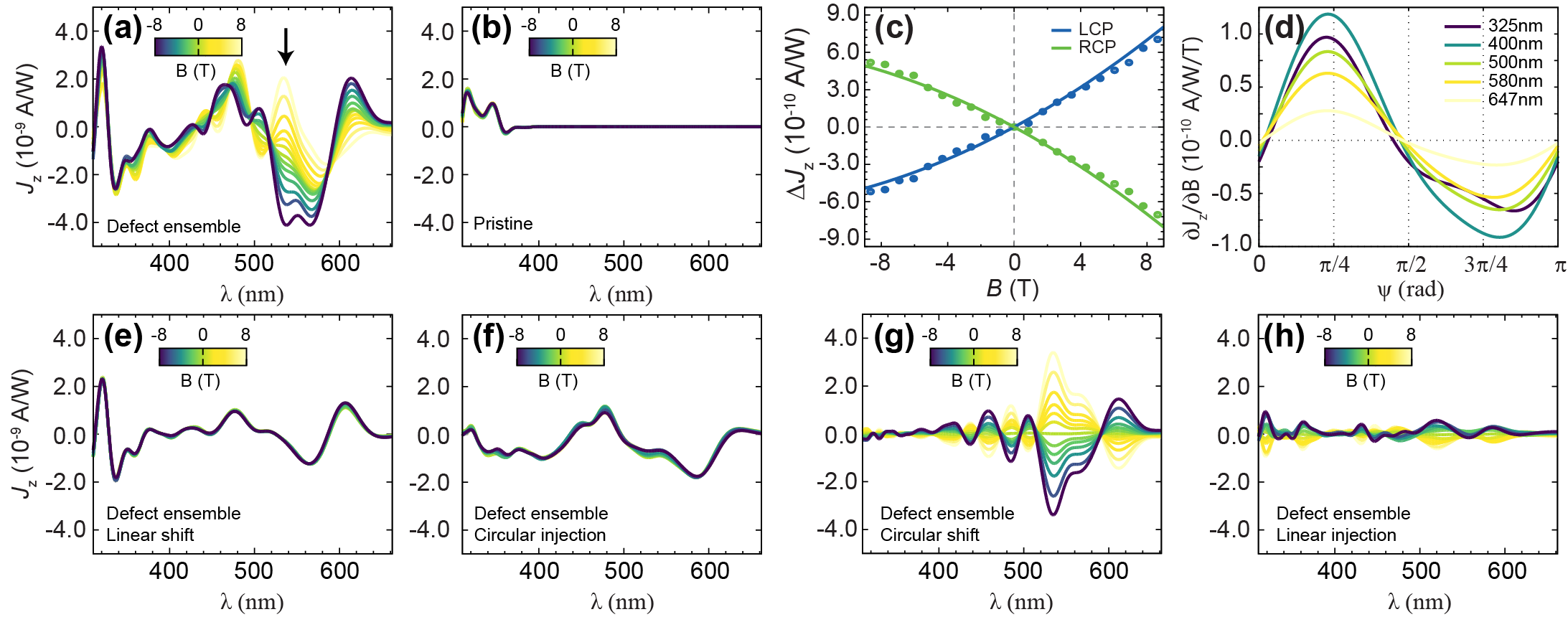}
    \caption{
    (a,b) Calculated spectral magneto-photocurrent of (a) defective BSO (defect ensemble) and (b) pristine BSO for $B\parallel[010]$ and $k\parallel[001]$.
    (c) Calculated magnetic field dependence of $\Delta J_z$ at $\lambda=568$~nm for quarter-wave angles $\psi\in[0,\pi]$ with fitting.
    (d) Magnetic sensitivity $\partial J_z/\partial B$ versus light-polarization state for all measured wavelengths fitting using Eq.~\ref{J_fitting}.
    (e-h) Decomposition photocurrent of the defect ensemble into (e) linear shift, (f) circular injection, (g) circular shift, and (h) linear injection currents. 
    }
    \label{fig3}
\end{figure*}

Having established that oxygen vacancies generate localized moments while preserving the macroscopic crystal symmetry at $B=0$, we now test whether field-selected symmetry lowering is sufficient to reproduce the three central experimental signatures simultaneously: a longitudinal odd-in-$B$ response, strong sub-band-gap photocurrent, and helicity-selective enhancement under circular polarization. We evaluate the resulting photocurrent using DFT with a perturbative formalism that includes both shift and injection contributions~\cite{sipe2000PRB,young2012first}. We include spin–orbit coupling and an external magnetic field as a Zeeman-like term applied along $[010]$. Longitudinal photocurrent with $xy$ oscillating electric field for the defect ensemble and for pristine BSO show significantly different behavior depending on magnetic field as shown in Fig. \ref{fig3}(a) and (b), respectively. In the pristine case, there is no magnetic field dependence. By contrast, the defect ensemble BSO shows a pronounced sub-gap photocurrent, and the $B$ dependence tracks the dominant experimental trends closely, supporting our analytic derivation above. Together with photocurrent calculations for other possible defects (See Fig.~\ref{figSI6}), these results identify defect-induced local moments as the origin of the symmetry-forbidden response observed experimentally.

To enable a quantitative comparison between theory and experiment, we examine the magnetic-field dependence of the photocurrent at the photon energy marked by the black arrow in Fig.~\ref{fig3}(a). Plotting $J_z$ versus $B$ reproduces the experimentally observed dominant linear-in-$B$, helicity-selective response [Fig.~\ref{fig3}(c)]. At the same field amplitude, the measured and calculated responsivities are $1.00 \times 10^{-10}$ A/W and $0.45 \times 10^{-10}$ A/W, respectively. We quantify magnetic sensitivity by extracting the slope $\partial J/\partial B$ for each $\psi$ from linear fits to the traces in Fig.~\ref{fig3}(d). The experimental sensitivity shows peaks for RCP $(\psi=\pi/4)$ and LCP $(\psi=3\pi/4)$ and reverses sign between the two across the visible range [Fig.~\ref{fig3}(c) and Fig.~\ref{figSI5}]. Consistently, fitting the numerical calculation to Eq.~\ref{J_fitting} yields a dominant circular component with much smaller linear terms, mirroring the experimental trend vs. wavelength. In all cases, they differ by about a factor of 2, and the residual amplitude discrepancy likely reflects differences in defect concentration or carrier lifetime, as well as limitations of the present supercell-based, perturbative treatment (lack of an explicit real-time vector potential $A(t)$ and omission of many-body/excitonic effects).

We now demonstrate the polarization dependence of the photocurrent. Because a magnetic field aligns pre-existing spins and lowers the symmetry, both linear and circular magneto-photocurrent tensors become allowed; however, the experimentally observed magnitude contrast [CPL $\approx 4\times$ LPL at $\lambda=568$~nm; Fig.~\ref{fig1}(f)] cannot be set by symmetry considerations alone. To resolve this, we decompose our BPVE (shift and injection current) into linear and circular components. From shift current ($J_{\rm{SC}}$) and injection current ($J_{\rm{IC}}$) equations \cite{wang2020electrically, zhang2019NatCommun},

\begin{equation}
\begin{aligned}
J^{a}_{\mathrm{SC}}
= &-\,\frac{i\pi e^{3}}{2\hbar^{2}}
\!\int\![d\mathbf{k}]\,
\sum_{mn\sigma}
f_{nm}\!\left(\,
r^{b}_{mn}\,r^{c}_{nm;k_{a}}-r^{c}_{mn}\,r^{b}_{nm;k_{a}}
\right) \\
&\delta(\omega_{nm}-\omega_{\beta})\,
E_{b}(\omega_{\beta})\,E_{c}(-\omega_{\beta}) \, .    
\end{aligned}
\label{SC}
\end{equation}

\begin{equation}
\begin{aligned}
\frac{d J^{a}_{\mathrm{IC}}}{dt}
= &-\,\frac{\pi e^{3}}{\hbar^{2}}
\!\int\![d\mathbf{k}]\,
\sum_{mn\sigma}
f_{mn}\,\Delta^{a}_{mn}\,
r^{b}_{nm}\,r^{c}_{mn}\,\\
&\delta(\omega_{nm}-\omega_{\beta})\,
E_{b}(\omega_{\beta})\,E_{c}(-\omega_{\beta}) \, .    
\end{aligned}
\label{IC}
\end{equation}
\noindent where \(r_{nm;k^a}^{\,b}=\frac{\partial r_{mn}^{\,b}}{\partial k^a} - i\,r_{mn}^{\,b}(A_{m}^{a}-A_{n}^{a})\) is the gauge-covariant derivative associated with the shift vector. \(r_{nm}^{\,b}=i\langle n|\partial_{k^b}|m \rangle \) and \(A_{n}^{a}=i\langle n|\partial_{k^a}|n \rangle \) are the interband and intraband Berry connections, respectively. \(\Delta^a_{nm}=v^a_{nn}-v^a_{mm}\), where \(v^a_{nm}=\frac{1}{\hbar}\langle n|\partial_{k^a}H|m \rangle \). We can symmetrize our electric fields as following;

\begin{equation}
\begin{aligned}
   E_{b}(\omega)\,E_{c}(-\omega)
 = E_{b}(\omega)\,E_{c}^{*}(\omega)
 = \operatorname{Re}\!\big(E_{b}E_{c}^{*}\big)
 + i\,\operatorname{Im}\!\big(E_{b}E_{c}^{*}\big)\, ,\\
 E_{b}(-\omega)\,E_{c}(\omega)
 = E_{b}^{*}(\omega)\,E_{c}(\omega)
 = \operatorname{Re}\!\big(E_{b}E_{c}^{*}\big)
 - i\,\operatorname{Im}\!\big(E_{b}E_{c}^{*}\big)\,.  
 \label{Efieldsymm}
\end{aligned}
\end{equation}
When we substitute Eq.~\ref{Efieldsymm} into Eqs.~\ref{SC} and ~\ref{IC}, we can decompose the linear and circular components as below;

\begin{equation}\label{SC_decom}
\begin{aligned}
&J^{a}_{\mathrm{SC}}=2\,\sigma^{abc}_{\mathrm{Ib}}\,\operatorname{Re}\!\big(E_{b}E_{c}^{*}\big) + 2i\,\sigma^{abc}_{\mathrm{TRb}}\,\operatorname{Im}\!\big(E_{b}E_{c}^{*}\big)\,,\\
&\sigma^{abc}_{\mathrm{Ib}}
\begin{aligned}
 = & -\,\frac{\pi e^{3}}{4\hbar^{2}}
\!\int\![d\mathbf{k}]\,
\sum_{mn\sigma}
f_{nm}\,
\big(R^{a;b}_{mn}(\mathbf{k})+R^{a;c}_{mn}(\mathbf{k})\big)\,\\
&\big(r^{b}_{nm}r^{c}_{mn}\big)\,
\delta(\omega_{nm}-\omega)\,, 
\end{aligned} \\
&\sigma^{abc}_{\mathrm{TRb}}
\begin{aligned}
= &-\,\frac{\pi e^{3}}{4i\,\hbar^{2}}
\!\int\![d\mathbf{k}]\,
\sum_{mn\sigma}
f_{nm}\,
\big(R^{a;b}_{mn}(\mathbf{k})+R^{a;c}_{mn}(\mathbf{k})\big)\, \\
&\Omega^{d}_{mn}(\mathbf{k})\,
\delta(\omega_{nm}-\omega)\,.    
\end{aligned}
\end{aligned}
\end{equation}

\begin{equation}\label{IC_decom}
\begin{aligned}
&\frac{d J^{a}_{\mathrm{IC}}}{dt}
= 2\,\eta^{abc}_{\mathrm{TRb}}\,
\operatorname{Re}\!\big(E_{b}E_{c}^{*}\big)
+ 2i\,\eta^{abc}_{\mathrm{Ib}}\,
\operatorname{Im}\!\big(E_{b}E_{c}^{*}\big)\,,\\
&\eta^{abc}_{\mathrm{Ib}}
= -\,\frac{\pi e^{3}}{2\hbar^{2}}
\!\int\![d\mathbf{k}]\,
\sum_{mn\sigma}
f_{mn}\,\Delta^{a}_{mn}\,
[r^b_{nm},r^c_{mn}],
\delta(\omega_{nm}-\omega)\,,\\
&\eta^{abc}_{\mathrm{TRb}}
= -\,\frac{\pi e^{3}}{2\hbar^{2}}
\!\int\![d\mathbf{k}]\,
\sum_{mn\sigma}
f_{mn}\,\Delta^{a}_{mn}\,
\{r^b_{nm},r^c_{mn}\},
\delta(\omega_{nm}-\omega)\,.
\end{aligned}
\end{equation}
Here, $R^{a;b}_{mn}(\mathbf{k})$ and $\Omega^{d}_{mn}(\mathbf{k})$ are shift vector and local Berry curvature, respectively. When $b \neq c$, $\operatorname{Re}\!\big(E_{b}E_{c}^{*}\big)$ and $\operatorname{Im}\!\big(E_{b}E_{c}^{*}\big)$ are the linearly and circularly polarized light, respectively. By their symmetry, $\sigma^{abc}_{\mathrm{Ib}}$ and $\eta^{abc}_{\mathrm{Ib}}$ are arise only when inversion symmetry breaking (Ib) occurs, and $\sigma^{abc}_{\mathrm{TRb}}$ and $\eta^{abc}_{\mathrm{TRb}}$ arise only when time-reversal symmetry breaking (TRb) occurs. We plot each component of Eqs.~\ref{SC_decom} and ~\ref{IC_decom} with respect to photon energy in Fig. \ref{fig3}(e)-(h). $\sigma^{abc}_{\mathrm{Ib}}$ and $\eta^{abc}_{\mathrm{Ib}}$ are non-zero due to non-centrosymmetry, but they have no dependence on magnetic field strength. $\sigma^{abc}_{\mathrm{TRb}}$ and $\eta^{abc}_{\mathrm{TRb}}$ are time-reversal symmetry broken response and therefore have magnetic field dependency as per symmetry analysis. We refer to $\sigma^{abc}_{\mathrm{TRb}}$ and $\eta^{abc}_{\mathrm{TRb}}$ as circular shift and linear injection respectively. We find that the circular shift component is around 10 times larger than the linear injection component at 568 nm, qualitatively consistent with our measurement, also showing in that the circular response dominates over the linear one.

\section{Discussion}

\begin{figure*}
    \centering
    \includegraphics[width=0.5\textwidth]{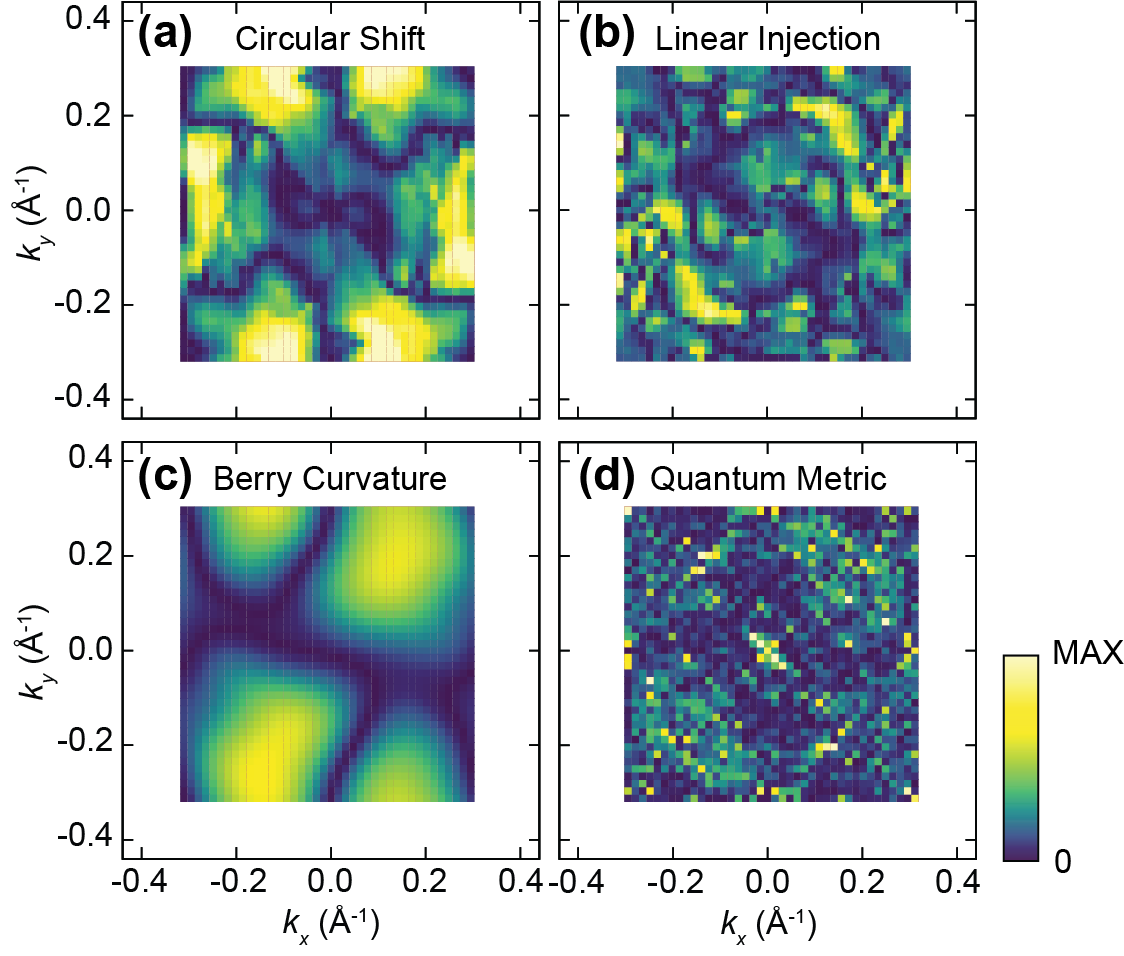}
    \caption{
    (a,b) Two-dimensional (2D) momentum-resolved distribution of (a) circular shift and (b) linear injection currents of the 568~nm peak in the first Brillouin zone (BZ) of the defect ensemble \ch{Bi_{12}SiO_{20}} at $k_z=0$ plane.
    (c,d) Corresponding 2D momentum-resolved distribution of (c) Berry curvature and (d) Quantum metric map in the first BZ of the defect ensemble \ch{Bi_{12}SiO_{20}} at $k_z=0$ plane.
    }
    \label{fig4}
\end{figure*}

To further elucidate the geometric origin of the defect-activated magneto-photocurrent, we examine the k-resolved distribution of the dominant time-reversal-breaking contributions and compare them directly with the underlying band-geometric quantities. As shown in Fig. \ref{fig4}(a) and \ref{fig4}(b), the circular shift and linear injection currents at 568 nm exhibit pronounced hot spots localized in specific regions of the Brillouin zone. Remarkably, these hot spots spatially correlate with the distributions of the band Berry curvature and quantum metric shown in Fig. \ref{fig4}(c) and \ref{fig4}(d), respectively. The circular shift component ($\Omega^{d}_{mn}(\mathbf{k})$) in Eq.~\ref{SC_decom} closely follows regions of enhanced Berry curvature, consistent with its dependence on the antisymmetric part of the interband matrix elements, while the linear injection contribution ($\{r^b_{nm},\,r^c_{mn}\}$) in Eq.~\ref{IC_decom} traces regions where the quantum metric is large, reflecting sensitivity to the symmetric part of the band-connection structure, as described in Supplementary Note  ~\ref{SI.NoteS3}.

This correspondence provides direct k-space evidence that the defect-enabled symmetry lowering does not merely lift selection rules but also activates latent band-geometric responses that are already encoded in the pristine electronic structure. The magnetic field selects a time-reversal-broken sector of the defect ensemble, thereby preventing cancellation of Berry-curvature- and quantum-metric-mediated contributions upon Brillouin-zone integration. Consequently, the observed helicity-selective longitudinal magneto-photocurrent can be understood as a field-unmasked geometric response rooted in the interplay among spin-orbit coupling, defect-localized moments, and band topology. 

\section{Conclusion}
\label{sec:conclusion}

We have demonstrated that nominally forbidden longitudinal magneto-photocurrents can be activated in chiral cubic Bi$_{12}$SiO$_{20}$ through defect-induced magnetic symmetry lowering. Polarization-resolved photocurrent spectroscopy reveals a pronounced odd-in-$B$, helicity-selective response in Voigt geometry across the visible range, persisting well below the band gap and thus pointing to the essential role of in-gap defect states. First-principles calculations identify oxygen vacancies as the microscopic origin; vacancy-bound electronic states generate localized spin moments on neighboring Bi-O units that are stabilized by strong spin-orbit coupling. While symmetry-related vacancy configurations remain nearly degenerate and preserve the macroscopic $T$ symmetry at $B=0$, an external magnetic field selects among spin-orbit-coupled anisotropy minima, introducing a unique axial direction and reducing the effective magnetic symmetry to the subgroup that leaves $\hat{\mathbf B}$ invariant, thereby lifting the longitudinal selection rule. The resulting photocurrent is dominated by time-reversal-breaking contributions and exhibits $k$-space hot spots that correlate closely with the band Berry curvature and quantum metric, establishing a direct connection between defect-unmasked symmetry channels and band geometry. Our results establish a general design principle: controlled defects combined with magnetic-field selection provide a route to helicity-tunable, symmetry-forbidden nonlinear magneto-optical functionalities in chiral quantum materials.

\section{Methods}
\subsection{Experimental details}

An Innova (Coherent) I-90 Ar/Kr ion laser and a Kimmon Koha IK5652R-G HeCd laser provided the incident laser illumination. Light polarization was modulated using a Pockels cell electro-optic  modulator (Conoptics KD*P-UV), operated at half-wave voltage. Following the modulator, and immediately before the sample, an achromatic quarter-wave plate was automatically rotated from $\psi=0^\circ$ to $\psi=180^\circ$ to vary the helicity of the incident modulated polarization from linear through elliptical to right circular polarization RCP ($\psi=45^\circ$), and left circular polarization LCP ($\psi=135^\circ$) states. The modulation frequency was 376.9 Hz, and data were recorded in phase with the polarization modulation. The modulation technique was reported in detail previously \cite{Burger:2019, Grachev:1982}. The BSO cubic crystal, 3\,$\times$\,3 mm, was mounted in a magneto-optic cryostation (Montana Instruments) and maintained under static vacuum conditions at room temperature. Photocurrents were recorded using an SRS-830 lock in amplifier. Quarter-wave plate rotation and data acquisition from the SR-830 were automated using LabVIEW. At $\psi = 0^\circ$, the incident light is linearly polarized along the [100] crystal direction at the entrance surface. Owing to the optically active of BSO, the polarization state evolves during propagation through the crystal, so that the local electric field generally acquires both $x$- and $y$-components inside the sample. Photocurrents were collected through indium tin oxide (ITO) transparent conducting electrodes deposited on the \([001]\) and \([00\overline{1}]\) crystal faces. The 100\,nm thick electrodes were sputtered at room temperature, following the methods described in Ref.~\cite{bennett2022bulk}. For each available laser wavelength, data were recorded during quarter-wave plate rotation, and measurements were acquired in sequence for each magnetic field strength.

The experimental configuration used to probe magneto-photocurrents is shown in Fig.~\ref{fig1}(b). A magnetic field is applied along $[010]$ while light propagates along $[001]$ ($k\parallel \hat{z}$). The incident polarization is controlled by an electro-optic modulator and a rotating quarter-wave plate (QWP), enabling continuous tuning from linear ($x$-polarized) to elliptical to circular; right-circular (RCP) and left-circular (LCP) states are realized at $\psi=45^\circ$ and $135^\circ$, respectively~\cite{Grachev:1982}. Magnetic fields up to $\pm 0.6$~T were applied in $0.1$~T steps, and the photocurrent was measured over 325–647 nm.

PL measurements were conducted with an excitation wavelength of 457 nm. A power of 20 mW was focused down to $\approx$ 15 micron spot size. The sample was cooled to 2 Kelvin before measurements.

\subsection{Theoretical details}
\subsubsection*{Computational framework}
First-principles calculations were performed within the framework of DFT using the projector-augmented wave (PAW) method~\cite{kresse1999PRB} as implemented in the \textsc{VASP} package~\cite{kresse1993PRB}. The exchange–correlation functional was treated within the generalized gradient approximation (GGA) of Perdew–Burke–Ernzerhof (PBE)~\cite{perdew1996PRL}. A plane-wave cutoff energy of 500~eV was employed, and the Brillouin zone was sampled with a $10 \times 10 \times 10$ grid, ensuring energy convergence within 1~meV/atom. Structural relaxations were performed until the residual Hellmann–Feynman forces on each atom were below 0.01~eV~\AA$^{-1}$. SOC was included self-consistently in all calculations using a noncollinear spinor Hamiltonian, because the strong relativistic effects of Bi are essential for accurately describing the magneto-optical response. In addition, we corrected the band-gap underestimation of DFT by applying a scissor operator, making a rigid conduction-band shift (experimental band gap : 3.2 eV and simulation band gap : 2.5 eV)~\cite{levine1989linear,nastos2005scissors}.  The effect of an external magnetic field ($B_\mathrm{ext}$) was incorporated through the Zeeman interaction added to the Kohn–Sham Hamiltonian as below.
\[
    \hat{H} = \hat{H}_0 + B_\mathrm{ext} \cdot \sigma
\]

\subsubsection*{Defect formation energy calculation for charged states}
To determine the charge state of the oxygen vacancy, we evaluate the defect formation energy as a function of the Fermi level.
For a defect $D$ in charge state $q$, the formation energy is
\begin{equation}
\begin{aligned}
  E_{\mathrm{form}}(D^{q};E_F)
  =
  & E_{\mathrm{tot}}(D^{q})
  -
  E_{\mathrm{tot}}(\mathrm{bulk})
  -
  \sum_i n_i\mu_i \\
  &+
  q\!\left(E_F+E_{\mathrm{VBM}}+\Delta V\right)
  +
  E_{\mathrm{corr}}(q),    
\end{aligned}
  \label{eq:defect_formation_energy}
\end{equation}
where $E_{\mathrm{tot}}(D^{q})$ and $E_{\mathrm{tot}}(\mathrm{bulk})$ are total energies of defective and pristine supercells;
$n_i$ and $\mu_i$ denote the number and chemical potential of species $i$ exchanged with reservoirs;
$E_{\mathrm{VBM}}$ is the valence-band maximum;
$\Delta V$ accounts for potential alignment; and $E_{\mathrm{corr}}(q)$ is the finite-size charge correction.

\subsubsection*{Nonlinear optical response calculations}

To evaluate the defect-activated magneto-photocurrent in \ch{Bi12SiO20}, we employed a perturbative second-order optical-response formalism that includes both shift-current and injection-current contributions~\cite{Young:2012, sipe2000PRB}. Starting from the Kohn-Sham Hamiltonian with the SOC and an external magnetic field introduced through a Zeeman term, we project the Kohn-Sham Hamiltonian to maximally localized Wannier function~\cite{mostofi2008CPC}. We computed the longitudinal photocurrent response where the Brillouin zone was sampled with a $25 \times 25 \times 25$ grid, which show sufficiently converged photocurrent spectra.

\section*{Acknowledgments}
This work was supported by the U.S. Department of Energy, Office of Science, Basic Energy Sciences, under Award No. DE-SC0024942. 

\bibliography{new_referece_260310}

@article{grinberg2013Nature,
  author  = {Grinberg, Ilya and West, D. Vincent and Torres, Maria and Gou, Gaoyang and Stein, David M. and Wu, Liyan and Chen, Guannan and Gallo, Eric M. and Akbashev, Andrew R. and Davies, Peter K. and others},
  title   = {Perovskite oxides for visible-light-absorbing ferroelectric and photovoltaic materials},
  journal = {Nature},
  volume  = {503},
  number  = {7477},
  pages   = {509--512},
  year    = {2013}
}

@article{mak2010PRL,
  author  = {Mak, Kin Fai and Lee, Changgu and Hone, James and Shan, Jie and Heinz, Tony F.},
  title   = {Atomically thin MoS$_2$: a new direct-gap semiconductor},
  journal = {Physical Review Letters},
  volume  = {105},
  number  = {13},
  pages   = {136805},
  year    = {2010}
}

@article{graham2013NatPhys,
  author  = {Graham, Matt W. and Shi, Su-Fei and Ralph, Daniel C. and Park, Jiwoong and McEuen, Paul L.},
  title   = {Photocurrent measurements of supercollision cooling in graphene},
  journal = {Nature Physics},
  volume  = {9},
  number  = {2},
  pages   = {103--108},
  year    = {2013}
}

@article{akamatsu2021Science,
  author  = {Akamatsu, Takatoshi and Ideue, Toshiya and Zhou, Ling and Dong, Yu and Kitamura, Sota and Yoshii, Mao and Yang, Dongyang and Onga, Masaru and Nakagawa, Yuji and Watanabe, Kenji and others},
  title   = {A van der Waals interface that creates in-plane polarization and a spontaneous photovoltaic effect},
  journal = {Science},
  volume  = {372},
  number  = {6537},
  pages   = {68--72},
  year    = {2021}
}

@article{zhang2019Nature,
  author  = {Zhang, Y. J. and Ideue, Toshiya and Onga, Masaru and Qin, Feng and Suzuki, Ryuji and Zak, Alla and Tenne, Reshef and Smet, J. H. and Iwasa, Yoshihiro},
  title   = {Enhanced intrinsic photovoltaic effect in tungsten disulfide nanotubes},
  journal = {Nature},
  volume  = {570},
  number  = {7761},
  pages   = {349--353},
  year    = {2019}
}

@article{ma2017NatPhys,
  author  = {Ma, Qiong and Xu, Su-Yang and Chan, Ching-Kit and Zhang, Cheng-Long and Chang, Guoqing and Lin, Yuxuan and Xie, Weiwei and Palacios, Tom{\'a}s and Lin, Hsin and Jia, Shuang and others},
  title   = {Direct optical detection of Weyl fermion chirality in a topological semimetal},
  journal = {Nature Physics},
  volume  = {13},
  number  = {9},
  pages   = {842--847},
  year    = {2017}
}

@article{ma2022Nature,
  author  = {Ma, Chao and Yuan, Shaofan and Cheung, Patrick and Watanabe, Kenji and Taniguchi, Takashi and Zhang, Fan and Xia, Fengnian},
  title   = {Intelligent infrared sensing enabled by tunable moir{\'e} quantum geometry},
  journal = {Nature},
  volume  = {604},
  number  = {7905},
  pages   = {266--272},
  year    = {2022}
}

@article{xu2018NatPhys,
  author  = {Xu, Su-Yang and Ma, Qiong and Shen, Huitao and Fatemi, Valla and Wu, Sanfeng and Chang, Tay-Rong and Chang, Guoqing and Valdivia, Andr{\'e}s M. Mier and Chan, Ching-Kit and Gibson, Quinn D. and others},
  title   = {Electrically switchable Berry curvature dipole in the monolayer topological insulator WTe$_2$},
  journal = {Nature Physics},
  volume  = {14},
  number  = {9},
  pages   = {900--906},
  year    = {2018}
}

@article{mciver2012NatNanotechnol,
  author  = {McIver, J. W. and Hsieh, David and Steinberg, Hadar and Jarillo-Herrero, Pablo and Gedik, Nuh},
  title   = {Control over topological insulator photocurrents with light polarization},
  journal = {Nature Nanotechnology},
  volume  = {7},
  number  = {2},
  pages   = {96--100},
  year    = {2012}
}

@article{yang2018AdvMater,
  author  = {Yang, Ming-Min and Alexe, Marin},
  title   = {Light-induced reversible control of ferroelectric polarization in BiFeO$_3$},
  journal = {Advanced Materials},
  volume  = {30},
  number  = {14},
  pages   = {1704908},
  year    = {2018}
}

@article{matsubara2022NatCommun,
  author  = {Matsubara, Masakazu and Kobayashi, Takatsugu and Watanabe, Hikaru and Yanase, Youichi and Iwata, Satoshi and Kato, Takeshi},
  title   = {Polarization-controlled tunable directional spin-driven photocurrents in a magnetic metamaterial with threefold rotational symmetry},
  journal = {Nature Communications},
  volume  = {13},
  number  = {1},
  pages   = {6708},
  year    = {2022}
}

@article{ji2019NatureMater,
  author  = {Ji, Zhurun and Liu, Gerui and Addison, Zachariah and Liu, Wenjing and Yu, Peng and Gao, Heng and Liu, Zheng and Rappe, Andrew M. and Kane, Charles L. and Mele, Eugene J. and others},
  title   = {Spatially dispersive circular photogalvanic effect in a Weyl semimetal},
  journal = {Nature Materials},
  volume  = {18},
  number  = {9},
  pages   = {955--962},
  year    = {2019}
}

@article{morimoto2016SciAdv,
  author  = {Morimoto, Takahiro and Nagaosa, Naoto},
  title   = {Topological nature of nonlinear optical effects in solids},
  journal = {Science Advances},
  volume  = {2},
  number  = {5},
  pages   = {e1501524},
  year    = {2016}
}

@article{ahn2020PRX,
  author  = {Ahn, Junyeong and Guo, Guang-Yu and Nagaosa, Naoto},
  title   = {Low-frequency divergence and quantum geometry of the bulk photovoltaic effect in topological semimetals},
  journal = {Physical Review X},
  volume  = {10},
  number  = {4},
  pages   = {041041},
  year    = {2020}
}

@article{ahn2022NatPhys,
  title={Riemannian geometry of resonant optical responses},
  author={Ahn, Junyeong and Guo, Guang-Yu and Nagaosa, Naoto and Vishwanath, Ashvin},
  journal={Nature Physics},
  volume={18},
  number={3},
  pages={290--295},
  year={2022},
  publisher={Nature Publishing Group UK London}
}

@article{zhang2019NatCommun,
  author  = {Zhang, Yang and Holder, Tobias and Ishizuka, Hiroaki and de Juan, Fernando and Nagaosa, Naoto and Felser, Claudia and Yan, Binghai},
  title   = {Switchable magnetic bulk photovoltaic effect in the two-dimensional magnet CrI$_3$},
  journal = {Nature Communications},
  volume  = {10},
  number  = {1},
  pages   = {3783},
  year    = {2019}
}

@article{ma2023NatRevPhys,
  author  = {Ma, Qiong and Krishna Kumar, Roshan and Xu, Su-Yang and Koppens, Frank H. L. and Song, Justin C. W.},
  title   = {Photocurrent as a multiphysics diagnostic of quantum materials},
  journal = {Nature Reviews Physics},
  volume  = {5},
  number  = {3},
  pages   = {170--184},
  year    = {2023}
}

@article{sirica2022NatMater,
  author  = {Sirica, Nicolas and Orth, Peter P. and Scheurer, Mathias S. and Dai, Y. M. and Lee, M.-C. and Padmanabhan, Prashant and Mix, L. T. and Teitelbaum, S. W. and Trigo, Mariano and Zhao, L. and others},
  title   = {Photocurrent-driven transient symmetry breaking in the Weyl semimetal TaAs},
  journal = {Nature Materials},
  volume  = {21},
  number  = {1},
  pages   = {62--66},
  year    = {2022}
}

@article{pettine2023Light,
  author  = {Pettine, Jacob and Padmanabhan, Prashant and Sirica, Nicholas and Prasankumar, Rohit P. and Taylor, Antoinette J. and Chen, Hou-Tong},
  title   = {Ultrafast terahertz emission from emerging symmetry-broken materials},
  journal = {Light: Science \& Applications},
  volume  = {12},
  number  = {1},
  pages   = {133},
  year    = {2023}
}

@article{Wang2024_NatComm_hiddenSpinCPC,
  author  = {Wang, Kexin and Zhang, Butian and Yan, Chengyu and Du, Luojun and Wang, Shun and others},
  title   = {Circular photocurrents in centrosymmetric semiconductors with hidden spin polarization},
  journal = {Nature Communications},
  volume  = {15},
  pages   = {9036},
  year    = {2024},
  doi     = {10.1038/s41467-024-53425-9}
}

@article{Cheon2023_npj2D_BPVE_PtSe2,
  author  = {Cheon, Cheol-Yeon and Sun, Zhe and Cao, Jiang and Gonzalez Marin, Juan Francisco and Tripathi, Mukesh and Watanabe, Kenji and Taniguchi, Takashi and Luisier, Mathieu and Kis, Andras},
  title   = {Disorder-induced bulk photovoltaic effect in a centrosymmetric van der Waals material},
  journal = {npj 2D Materials and Applications},
  volume  = {7},
  pages   = {74},
  year    = {2023},
  doi     = {10.1038/s41699-023-00435-8}
}

@article{koch1976Ferro,
  author  = {Koch, W. T. H. and Munser, R. and Ruppel, W. and W{\"u}rfel, P.},
  title   = {Anomalous photovoltage in BaTiO$_3$},
  journal = {Ferroelectrics},
  volume  = {13},
  number  = {1},
  pages   = {305--307},
  year    = {1976}
}

@article{Young:2012,
  author  = {Young, Steve M. and Rappe, Andrew M.},
  title   = {First principles calculation of the shift current photovoltaic effect in ferroelectrics},
  journal = {Physical Review Letters},
  volume  = {109},
  number  = {11},
  pages   = {116601},
  year    = {2012}
}

@article{spanier2016NatPhoton,
  author  = {Spanier, Jonathan E. and Fridkin, Vladimir M. and Rappe, Andrew M. and Akbashev, Andrew R. and Polemi, Alessia and Qi, Yubo and Gu, Zongquan and Young, Steve M. and Hawley, Christopher J. and Imbrenda, Dominic and others},
  title   = {Power conversion efficiency exceeding the Shockley-Queisser limit in a ferroelectric insulator},
  journal = {Nature Photonics},
  volume  = {10},
  number  = {9},
  pages   = {611--616},
  year    = {2016}
}

@article{aftab2022AdvOptMater,
  author  = {Aftab, Sikandar and Iqbal, Muhammad Zahir and Haider, Zeeshan and Iqbal, Muhammad Waqas and Nazir, Ghazanfar and Shehzad, Muhammad Arslan},
  title   = {Bulk photovoltaic effect in 2D materials for solar-power harvesting},
  journal = {Advanced Optical Materials},
  volume  = {10},
  number  = {23},
  pages   = {2201288},
  year    = {2022}
}

@article{Burger:2019,
  author  = {Burger, Aaron M. and Agarwal, Radhe and Aprelev, Alexey and Schruba, Edward and Gutierrez-Perez, Alejandro and Fridkin, Vladimir M. and Spanier, Jonathan E.},
  title   = {Direct observation of shift and ballistic photovoltaic currents},
  journal = {Science Advances},
  volume  = {5},
  number  = {1},
  pages   = {eaau5588},
  year    = {2019}
}

@article{Burger:2020,
  author  = {Burger, Aaron M. and Gao, Lingyuan and Agarwal, Radhe and Aprelev, Alexey and Spanier, Jonathan E. and Rappe, Andrew M. and Fridkin, Vladimir M.},
  title   = {Shift photovoltaic current and magnetically induced bulk photocurrent in piezoelectric sillenite crystals},
  journal = {Physical Review B},
  volume  = {102},
  number  = {8},
  pages   = {081113},
  year    = {2020}
}

@article{dai2023PRB,
  author  = {Dai, Zhenbang and Rappe, Andrew M.},
  title   = {Magnetic bulk photovoltaic effect: Strong and weak field},
  journal = {Physical Review B},
  volume  = {107},
  number  = {20},
  pages   = {L201201},
  year    = {2023}
}

@article{pi2023magnetic,
  author  = {Pi, Hanqi and Zhang, Shuai and Weng, Hongming},
  title   = {Magnetic bulk photovoltaic effect as a probe of magnetic structures of EuSn$_2$As$_2$},
  journal = {Quantum Frontiers},
  volume  = {2},
  number  = {1},
  pages   = {6},
  year    = {2023}
}

@article{fridkin1985magnetophotovoltaic,
  author  = {Fridkin, V. M. and Lazarev, V. G. and Shlensky, A. L.},
  title   = {The magnetophotovoltaic effect in the crystals without a centre of symmetry},
  journal = {Japanese Journal of Applied Physics},
  volume  = {24},
  number  = {S2},
  pages   = {296},
  year    = {1985}
}

@article{Tassev:2000,
  title={Faraday effect of BSO and BTO crystals doped with Cr, Mn and Cu},
  author={Tassev, V and Gospodinov, M and Veleva, M},
  journal={Crystal Research and Technology: Journal of Experimental and Industrial Crystallography},
  volume={35},
  number={2},
  pages={213--219},
  year={2000},
  publisher={Wiley Online Library}
}

@article{Mihailovic:2008,
  author  = {Mihailovic, Pedja and Petricevic, Slobodan and Stankovic, Stevan and Radunovic, Jovan},
  title   = {Temperature dependence of the Bi$_{12}$GeO$_{20}$ optical activity},
  journal = {Optical Materials},
  volume  = {30},
  number  = {7},
  pages   = {1079--1082},
  year    = {2008},
  doi     = {10.1016/J.OPTMAT.2007.05.014}
}

@article{Burkov:2001,
  author  = {Burkov, V. I. and Egorysheva, A. V. and Kargin, Y. F.},
  title   = {Optical and chiro-optical properties of crystals with sillenite structure},
  journal = {Crystallography Reports},
  volume  = {46},
  number  = {2},
  pages   = {312--335},
  year    = {2001},
  doi     = {10.1134/1.1358415}
}

@article{ahmad2009,
  author  = {Ahmad, Ijaz and Marinova, Vera and Goovaerts, Etienne},
  title   = {High-frequency electron paramagnetic resonance of the hole-trapped antisite bismuth center in photorefractive bismuth sillenite crystals},
  journal = {Physical Review B},
  volume  = {79},
  number  = {3},
  pages   = {033107},
  year    = {2009}
}

@article{Reyher:1992,
  title={Optically detected magnetic resonance of the bismuth-on-metal-site intrinsic defect in photorefractive sillenite crystals},
  author={Reyher, H-J and Hellwig, U and Thiemann, O},
  journal={Physical Review B},
  volume={47},
  number={10},
  pages={5638},
  year={1993},
  publisher={APS}
}

@article{Ivchenko:2017,
  author  = {Ivchenko, E. L. and Ganichev, S. D.},
  title   = {Spin-dependent photogalvanic effects},
  journal = {arXiv preprint arXiv:1710.09223},
  year    = {2017}
}

@article{Hornung:2021,
  author  = {Hornung, Dieter and von Baltz, Ralph},
  title   = {Quantum kinetics of the magnetophotogalvanic effect},
  journal = {Physical Review B},
  volume  = {103},
  number  = {19},
  pages   = {195203},
  year    = {2021}
}

@article{petrov1979photogalvanic,
  author  = {Petrov, M. P. and Grachev, A. I.},
  title   = {Photogalvanic effects in bismuth silicate (Bi$_{12}$SiO$_{20}$)},
  journal = {ZhETF Pisma Redaktsiiu},
  volume  = {30},
  pages   = {18--21},
  year    = {1979}
}

@article{Grachev:1982,
  author  = {Grachev, A. L. and Petrov, M. P.},
  title   = {Photogalvanic effects in bismuth oxide compounds in the impurity absorption region},
  journal = {Ferroelectrics},
  volume  = {43},
  number  = {1},
  pages   = {181--184},
  year    = {1982}
}

@article{hou1973transport,
  author  = {Hou, S. L. and Lauer, R. B. and Aldrich, R. E.},
  title   = {Transport processes of photoinduced carriers in Bi$_{12}$SiO$_{20}$},
  journal = {Journal of Applied Physics},
  volume  = {44},
  number  = {6},
  pages   = {2652--2658},
  year    = {1973}
}

@article{isik2021defect,
  author  = {Isik, M. and Delice, S. and Gasanly, N. M.},
  title   = {Investigation of defect levels in Bi$_{12}$SiO$_{20}$ single crystals by thermally stimulated current measurements},
  journal = {Physica Scripta},
  volume  = {96},
  number  = {12},
  pages   = {125875},
  year    = {2021},
  doi     = {10.1088/1402-4896/ac4190}
}

@article{Lima:2011,
  author  = {Lima, A. F. and Farias, S. A. S. and Lalic, M. V.},
  title   = {Structural, electronic, optical, and magneto-optical properties of Bi$_{12}$MO$_{20}$ (M = Ti, Ge, Si) sillenite crystals from first principles calculations},
  journal = {Journal of Applied Physics},
  volume  = {110},
  number  = {8},
  pages   = {083705},
  year    = {2011},
  doi     = {10.1063/1.3652751}
}

@article{lima2013first,
  author  = {Lima, A. F. and Lalic, M. V.},
  title   = {First-principles study of the BiMO$_4$ antisite defect in the Bi$_{12}$MO$_{20}$ (M = Si, Ge, Ti) sillenite compounds},
  journal = {Journal of Physics: Condensed Matter},
  volume  = {25},
  number  = {49},
  pages   = {495505},
  year    = {2013}
}

@article{lauer1971thermally,
  author  = {Lauer, R. B.},
  title   = {Thermally stimulated currents and luminescence in Bi$_{12}$SiO$_{20}$ and Bi$_{12}$GeO$_{20}$},
  journal = {Journal of Applied Physics},
  volume  = {42},
  number  = {5},
  pages   = {2147--2149},
  year    = {1971}
}

@article{nair2012spin,
  author  = {Nair, R. R. and Sepioni, M. and Tsai, I.-Ling and Lehtinen, O. and Keinonen, J. and Krasheninnikov, Arkady V. and Thomson, T. and Geim, A. K. and Grigorieva, I. V.},
  title   = {Spin-half paramagnetism in graphene induced by point defects},
  journal = {Nature Physics},
  volume  = {8},
  number  = {3},
  pages   = {199--202},
  year    = {2012}
}

@article{yazyev2010emergence,
  author  = {Yazyev, Oleg V.},
  title   = {Emergence of magnetism in graphene materials and nanostructures},
  journal = {Reports on Progress in Physics},
  volume  = {73},
  number  = {5},
  pages   = {056501},
  year    = {2010}
}

@article{brand2025APR,
  title={Defect-induced magnetic symmetry breaking in oxide materials},
  author={Brand, Eric and Rosendal, Victor and Wu, Yichen and Tran, Thomas and Palliotto, Alessandro and Maznichenko, Igor V and Ostanin, Sergey and Esposito, Vincenzo and Ernst, Arthur and Zhou, Shengqiang and others},
  journal={Applied Physics Reviews},
  volume={12},
  number={1},
  pages={011327},
  year={2025},
  publisher={AIP Publishing}
}

@phdthesis{Sekhar2012BSOThesis,
  author  = {Sekhar, H.},
  title   = {Preparation, characterization and nonlinear optical absorption studies in Bi$_{12}$SiO$_{20}$, Cu$_2$O and CdS nanomaterials},
  school  = {School of Physics, University of Hyderabad},
  address = {Hyderabad, India},
  year    = {2012}
}

@article{sipe2000PRB,
  author  = {Sipe, J. E. and Shkrebtii, A. I.},
  title   = {Second-order optical response in semiconductors},
  journal = {Physical Review B},
  volume  = {61},
  number  = {8},
  pages   = {5337},
  year    = {2000}
}

@article{young2012first,
  author  = {Young, Steve M. and Zheng, Fan and Rappe, Andrew M.},
  title   = {First-principles calculation of the bulk photovoltaic effect in bismuth ferrite},
  journal = {Physical Review Letters},
  volume  = {109},
  number  = {23},
  pages   = {236601},
  year    = {2012}
}

@article{wang2020electrically,
  author  = {Wang, Hua and Qian, Xiaofeng},
  title   = {Electrically and magnetically switchable nonlinear photocurrent in PT-symmetric magnetic topological quantum materials},
  journal = {npj Computational Materials},
  volume  = {6},
  number  = {1},
  pages   = {199},
  year    = {2020}
}

@phdthesis{bennett2022bulk,
  author  = {Bennett-Jackson, Andrew L.},
  title   = {The Bulk Photovoltaic and Space-Charge Effects in Epitaxial BaTiO$_3$ Thin Films},
  school  = {Drexel University},
  year    = {2022}
}

@article{levine1989linear,
  author  = {Levine, Zachary H. and Allan, Douglas C.},
  title   = {Linear optical response in silicon and germanium including self-energy effects},
  journal = {Physical Review Letters},
  volume  = {63},
  number  = {16},
  pages   = {1719},
  year    = {1989}
}

@article{nastos2005scissors,
  author  = {Nastos, Fred and Olejnik, Bernd and Schwarz, Karlheinz and Sipe, J. E.},
  title   = {Scissors implementation within length-gauge formulations of the frequency-dependent nonlinear optical response of semiconductors},
  journal = {Physical Review B},
  volume  = {72},
  number  = {4},
  pages   = {045223},
  year    = {2005}
}

@article{mostofi2008CPC,
  author  = {Mostofi, Arash A. and Yates, Jonathan R. and Lee, Young-Su and Souza, Ivo and Vanderbilt, David and Marzari, Nicola},
  title   = {wannier90: A tool for obtaining maximally localised Wannier functions},
  journal = {Computer Physics Communications},
  volume  = {178},
  number  = {9},
  pages   = {685--699},
  year    = {2008}
}

@misc{sturman1992photovoltaic,
  author    = {Sturman, B. I. and Fridkin, V. M.},
  title     = {Photovoltaic Effect in Media without a Center of Symmetry and Related Phenomena},
  publisher = {Nauka},
  address   = {Moscow},
  year      = {1992}
}

@article{kresse1999PRB,
  title={From ultrasoft pseudopotentials to the projector augmented-wave method},
  author={Kresse, Georg and Joubert, Daniel},
  journal={Physical Review B},
  volume={59},
  number={3},
  pages={1758},
  year={1999},
  publisher={APS}
}

@article{kresse1993PRB,
  title={Ab initio molecular dynamics for liquid metals},
  author={Kresse, Georg and Hafner, J{\"u}rgen},
  journal={Physical Review B},
  volume={47},
  number={1},
  pages={558},
  year={1993},
  publisher={APS}
}

@article{perdew1996PRL,
  title={Generalized gradient approximation made simple},
  author={Perdew, John P and Burke, Kieron and Ernzerhof, Matthias},
  journal={Physical Review Letters},
  volume={77},
  number={18},
  pages={3865},
  year={1996},
  publisher={APS}
}

@article{wojdowski1985vibrational,
  title={Vibrational modes in {$Bi_{12}GeO_{20}$} and {$Bi_{12}SiO_{20}$} crystals},
  author={Wojdowski, W},
  journal={Physica Status Solidi (b)},
  volume={130},
  number={1},
  pages={121--130},
  year={1985},
  publisher={Wiley Online Library}
}

@article{lazarevic2013raman,
  title={Raman spectroscopy of bismuth silicon oxide single crystals grown by the {Czochralski} technique},
  author={Lazarevi{\'c}, Z and Kosti{\'c}, S and Radojevi{\'c}, Vesna and Rom{\v{c}}evi{\'c}, M and Gili{\'c}, Martina and Petrovi{\'c}-Damjanovi{\'c}, M and Rom{\v{c}}evi{\'c}, N},
  journal={Physica Scripta},
  volume={157},
  number={1},
  pages={014046},
  year={2013},
  publisher={IOP Publishing}
}

@article{isik2019temperature,
  title={Temperature-dependent band gap characteristics of {$Bi_{12}SiO_{20}$} single crystals},
  author={Isik, M and Delice, S and Gasanly, NM and Darvishov, NH and Bagiev, VE},
  journal={Journal of Applied Physics},
  volume={126},
  number={24},
  year={2019},
  publisher={AIP Publishing}
}

@article{cremades1996cathodoluminescence,
  title={Cathodoluminescence and photoluminescence in the core region of {$Bi_{12}GeO_{20}$} and {$Bi_{12}SiO_{20}$} crystals},
  author={Cremades, A and Santos, MT and Rem{\'o}n, A and Garc{\'\i}a, JA and Di{\'e}guez, E and Piqueras, J},
  journal={Journal of Applied Physics},
  volume={79},
  number={9},
  pages={7186--7190},
  year={1996},
  publisher={American Institute of Physics}
}

@article{Ma2021,
  author = {Ma, Qiong and Grushin, Adolfo G. and Burch, Kenneth S.},
  title = {Topology and geometry under the nonlinear electromagnetic spotlight},
  journal = {Nature Materials},
  volume = {20},
  pages = {1601-1614},
  year = {2021},
  doi = {10.1038/s41563-021-00992-7}
}

@article{Kang2025,
  author = {Kang, Mingu and Kim, Sunje and Qian, Yuting and Neves, Paul M. and Ye, Linda and Jung, Junseo and Puntel, Denny and Mazzola, Federico and Fang, Shiang and Jozwiak, Chris and Bostwick, Aaron and Rotenberg, Eli and Fuji, Jun and Vobornik, Ivana and Park, Jae-Hoon and Checkelsky, Joseph G. and Yang, Bohm-Jung and Comin, Riccardo},
  title = {Measurements of the quantum geometric tensor in solids},
  journal = {Nature Physics},
  volume = {21},
  pages = {110-117},
  year = {2025},
  doi = {10.1038/s41567-024-02678-8}
}

@article{krishna2025NatMater,
  title={Terahertz photocurrent probe of quantum geometry and interactions in magic-angle twisted bilayer graphene},
  author={Krishna Kumar, Roshan and Li, Geng and Bertini, Riccardo and Chaudhary, Swati and Nowakowski, Krystian and Park, Jeong Min and Castilla, Sebastian and Zhan, Zhen and Pantale{\'o}n, Pierre A and Agarwal, Hitesh and others},
  journal={Nature materials},
  volume={24},
  number={7},
  pages={1034--1041},
  year={2025},
  publisher={Nature Publishing Group UK London}
}
\newpage

\newpage

\newpage

\newpage

\clearpage
\newpage
\widetext  
\setcounter{equation}{0}
\setcounter{figure}{0}
\setcounter{table}{0}
\renewcommand{\theequation}{S\arabic{equation}}
\renewcommand{\thefigure}{S\arabic{figure}}

\begin{center}
\textbf{\large Extended data Figure 1}    
\end{center}

\begin{figure}[ht]
    \centering
    \includegraphics[width=.9\textwidth]{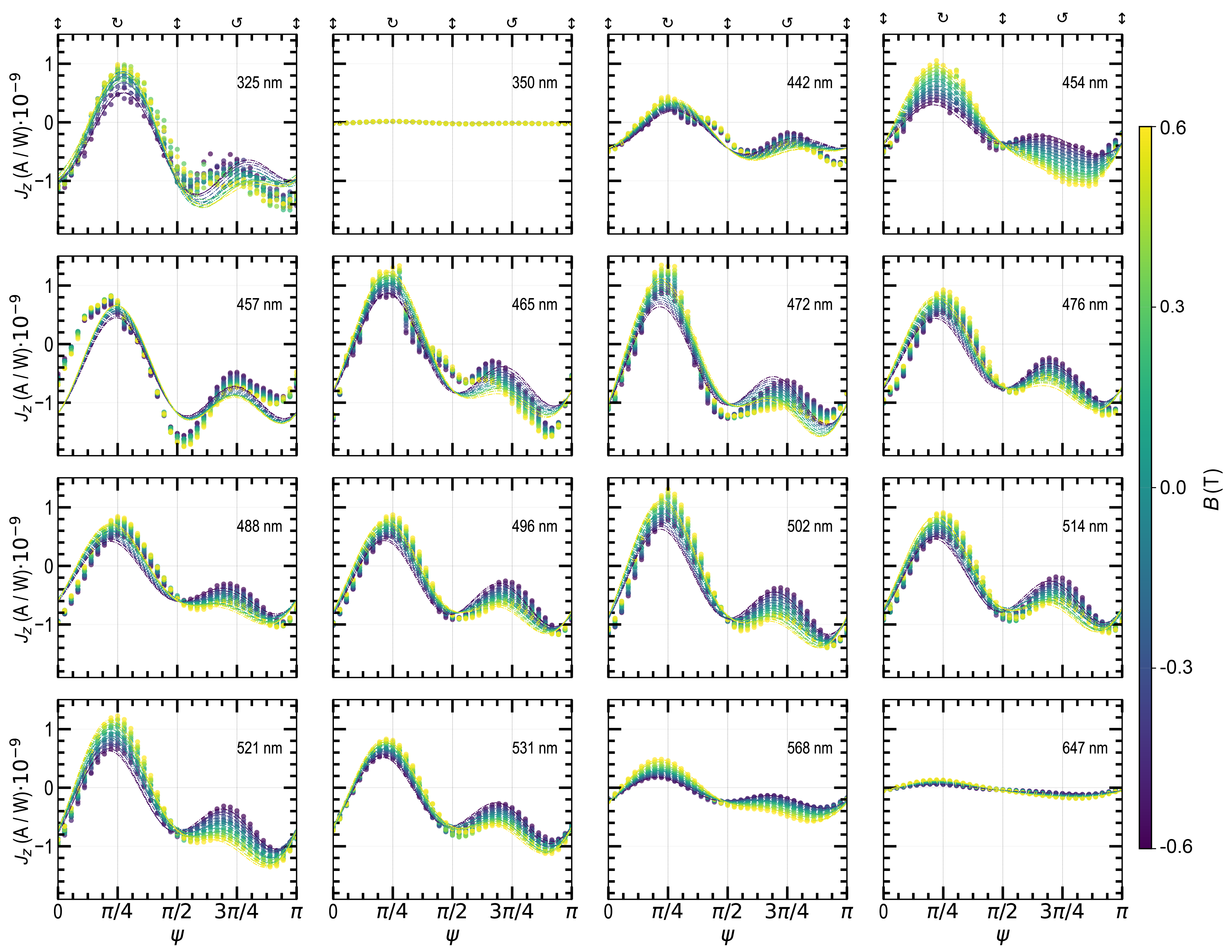}
    \caption{Measured variation in the longitudinal photocurrent vs. QWP angle for various magnetic fields. Individual traces fit using eq. \ref{J_fitting}.}
    \label{figSI3}
\end{figure}

The photocurrent was analyzed by fitting $J_z(\psi,B)$ at each wavelength across all magnetic fields to
$$
J_z(\psi, B) = J_L(B)\sin(4\psi+\psi_0) - J_C(B)\sin(2\psi) + J_0,
$$
where the amplitudes were taken to vary linearly with field, $J_L(B) = J_{L0} + J_{L1} B$ and $J_C(B) = J_{C0} + J_{C1} B$. The phase offset $\psi_0$ and constant offset $J_0$ were treated as field-independent and shared across all magnetic fields at a given wavelength. More general models allowing additional field-dependent terms were considered but did not improve the fit, and were therefore excluded. In some cases, the linear current $J_L$ exhibited no measurable field dependence, and was well described by $J_L(B)=J_{L}(B=0)$. This procedure yields a minimal description of the data that isolates the field dependence of the linear and circular photocurrent components.

\newpage
\begin{center}
\textbf{\large Extended data Figure 2}    
\end{center}

\begin{figure}[ht]
    \centering
    \includegraphics[width=.9\textwidth]{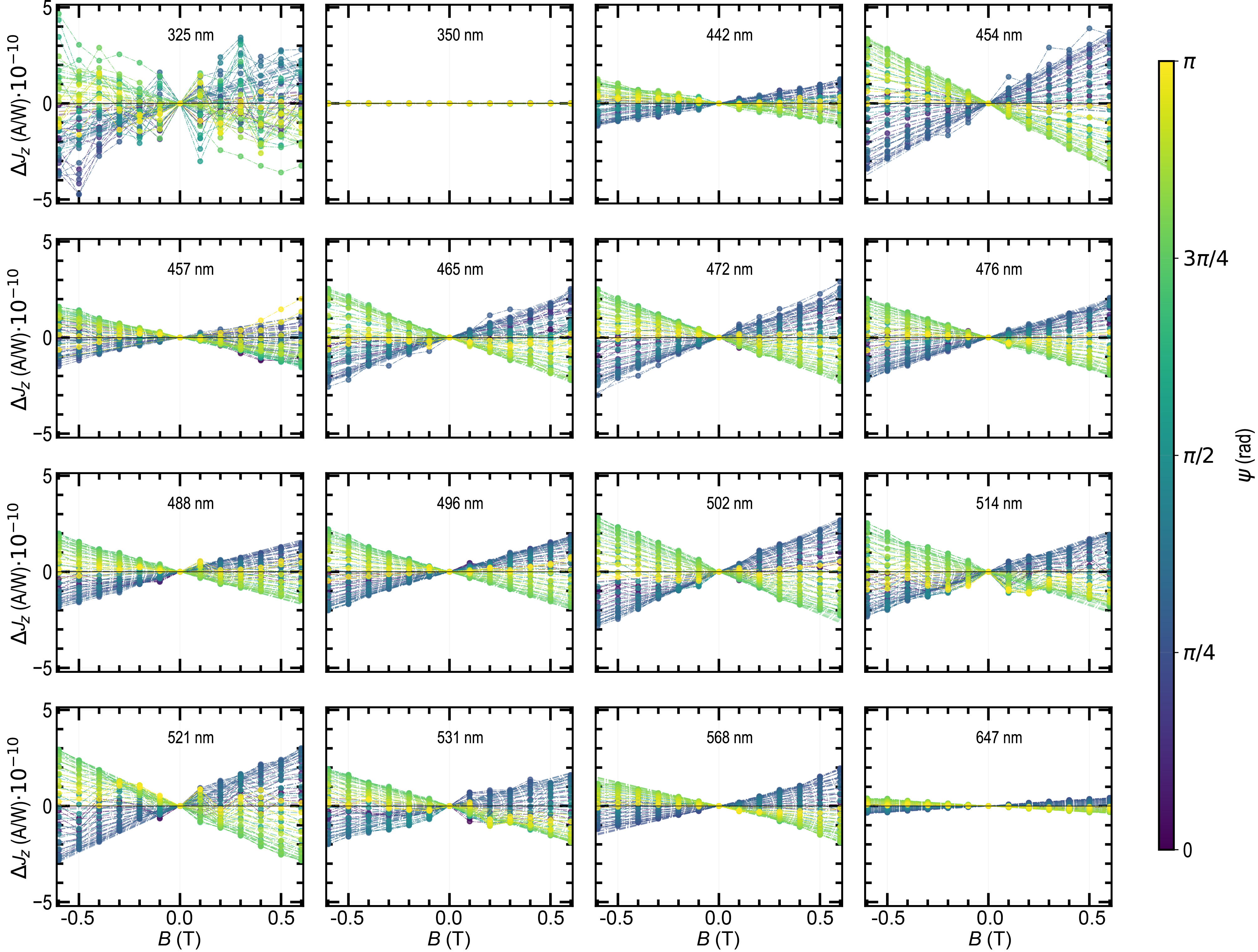}
    \caption{Measured magneto-optical photocurrent versus magnetic field for a range of QWP helicities at each laser wavelength excitation.}
    \label{figSI4}
\end{figure}

    

\newpage
\begin{center} 
\textbf{\large Extended data Figure 3}    
\end{center}

\begin{figure}[ht]
    \centering
    \includegraphics[width=.9\textwidth]{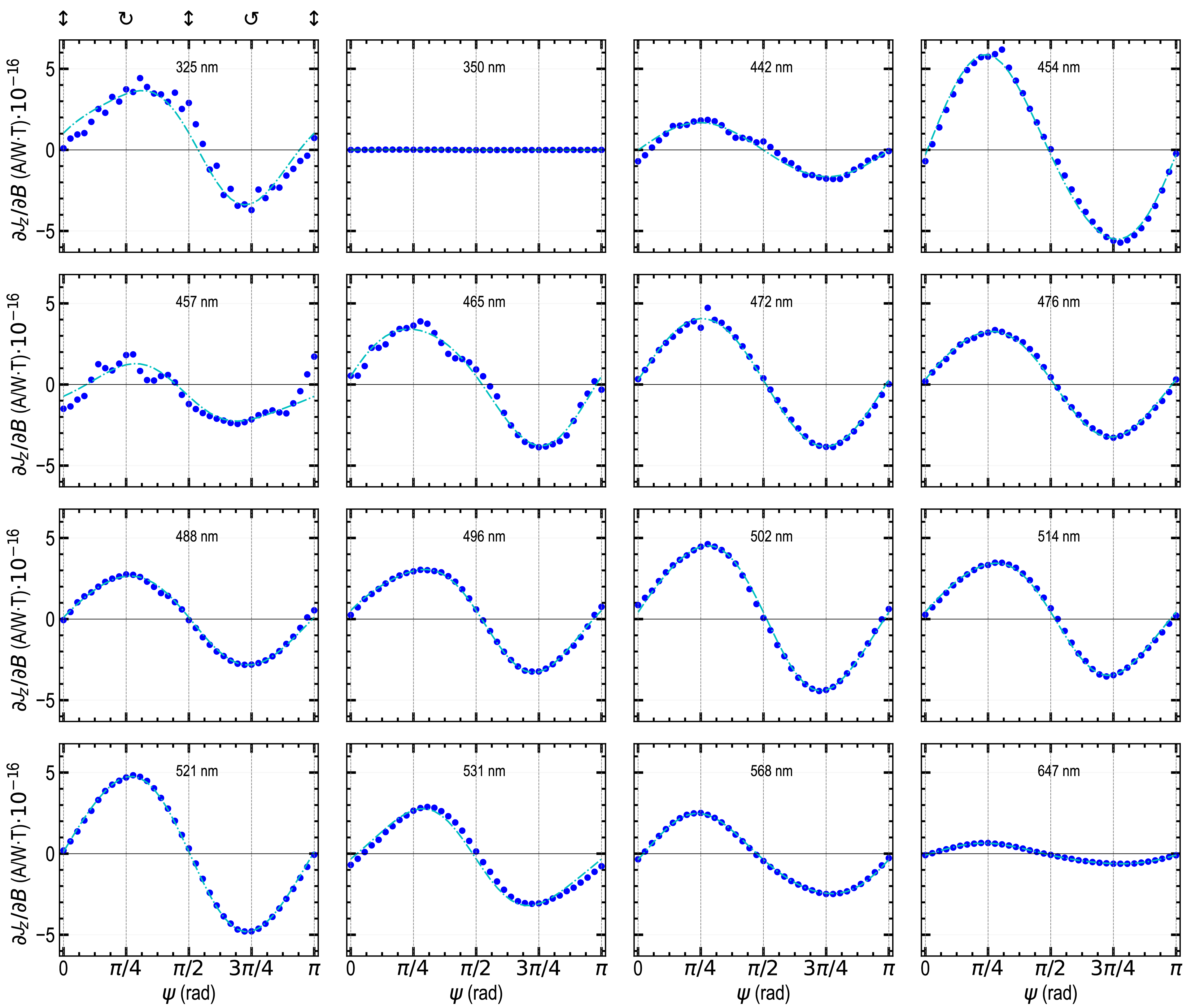}
    \caption{Light polarization variation of the photocurrent magnetic field sensitivity for each of the laser wavelengths.}
    \label{figSI5}
\end{figure}

\newpage
\begin{center}
\textbf{\large Extended data Figure 4}      
\end{center}

\begin{figure}[ht]
    \centering
    \includegraphics[width=.7\textwidth]{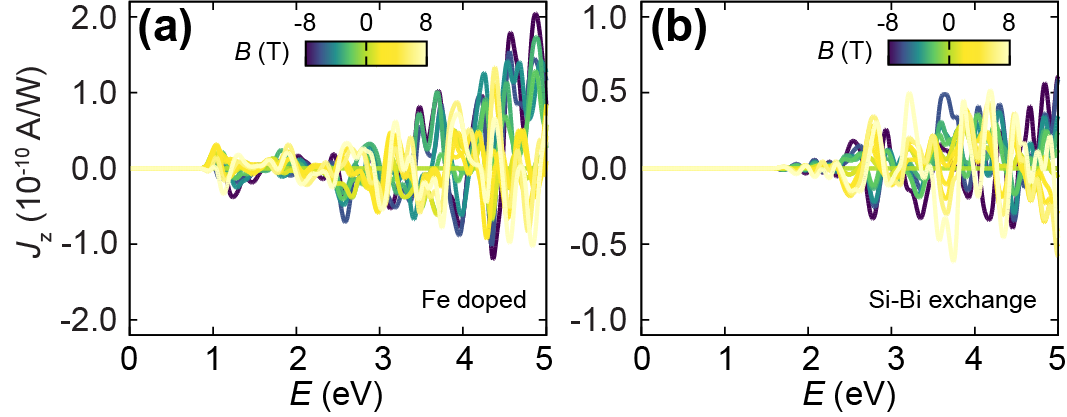}
    \caption{(a,b) Spectral magneto-photocurrent of (a) Fe-doped BSO and (b) Si-Bi exchange defective BSO defined as $\Delta j_z(B)\equiv j_z(B)-j_z(0)$ for $B\parallel[010]$ and $k\parallel[001]$. Both defect ensemble show magnetic field dependence, but not clear linear dependence on $B$ field.}
    \label{figSI6}
\end{figure}

\newpage
\clearpage
\setcounter{section}{0}
\renewcommand{\thesection}{S\arabic{section}}
\renewcommand{\thesubsection}{\Alph{subsection}}

\refstepcounter{section}
\section*{Note \thesection. Symmetry proof for the defect-activated longitudinal response}
\label{SI.NoteS1}

In this note, we summarize the symmetry logic underlying the main-text claim that a longitudinal odd-in-$B$ magneto-photocurrent is forbidden in pristine \ch{Bi12SiO20} with point group $T$, but can become allowed once a magnetic field selects a reduced symmetry sector of the defect ensemble.

\subsection{Pristine $T$ crystal: allowed second-order response without magnetic field}
\label{SI.NoteS1.A}

The point group $T$ is noncentrosymmetric and contains only proper rotations: the identity, eight $C_3$ rotations, and three $C_2$ rotations. In the absence of a magnetic field, the second-order photocurrent can be written as
\begin{equation}
    J_a = S_{abc} E_b E_c^{*} + i\,Q_{ab}\,(\mathbf{E} \times \mathbf{E}^{*})_{b},
\end{equation}
where $S_{abc}$ and $Q_{ab}$ denote the linear and circular photovoltaic tensors, respectively.

For light propagating along $\hat{\mathbf z}$ with polarization in the $xy$ plane, we use
\begin{equation}
    \mathbf{E}_l=\frac{E_0}{\sqrt{2}} (1,1,0)\, e^{i(kz-\omega t)},\qquad
    \mathbf{E}_c=\frac{E_0}{\sqrt{2}} (1,\pm i,0)\, e^{i(kz-\omega t)},
\end{equation}
where the sign distinguishes the two opposite helicities. The longitudinal response takes the form
\begin{equation}
    \mathbf{J} \cdot \hat{\mathbf z} =
    S_{zxy}\, (\mathbf{E}_l \cdot \hat{\mathbf x})(\mathbf{E}_l^{*} \cdot \hat{\mathbf y})
    + i\,Q_{zz}\,\big[(\mathbf{E}_c \times \mathbf{E}_c^{*}) \cdot \hat{\mathbf z}\big].
    \label{longiphoto}
\end{equation}
Under any of the three $C_2$ rotations of $T$, the signs of the relevant vector components transform in such a way that \cref{longiphoto} remains symmetry-allowed. Thus, in pristine point group $T$, the longitudinal zero-field LPGE and CPGE channels are not forbidden by symmetry.

\subsection{Pristine $T$ crystal in Voigt geometry: forbidden odd-in-$B$ longitudinal response}
\label{SI.NoteS1.B}

In the Voigt geometry used in the experiment, with $\mathbf{k}\parallel \hat{\mathbf z}$ and $\mathbf{B}\parallel \hat{\mathbf y}$, the longitudinal odd-in-$B$ magneto-photocurrent can be written as
\begin{equation}
    \mathbf{J} \cdot \hat{\mathbf z} =
    S_{zxyy}\, (\mathbf{E}_l \cdot \hat{\mathbf x})(\mathbf{E}_l^{*} \cdot \hat{\mathbf y})(\mathbf{B} \cdot \hat{\mathbf y})
    + i\,Q_{zzy}\,\big[(\mathbf{E}_c \times \mathbf{E}_c^{*}) \cdot \hat{\mathbf z}\big](\mathbf{B}\cdot \hat{\mathbf y}).
    \label{longimagnetophoto}
\end{equation}

The crucial point is that, in the pristine $T$ crystal, the two $C_2$ operations orthogonal to the magnetic-field direction still constrain the response. Under $C_{2x}$, both $(\mathbf{E}_l^{*}\cdot\hat{\mathbf y})$ and $(\mathbf{B}\cdot\hat{\mathbf y})$ change sign, while $\mathbf{J}\cdot\hat{\mathbf z}$ also changes sign. Comparing the transformed expression with \cref{longimagnetophoto} shows that the tensor coefficients associated with the odd-in-$B$ longitudinal response must vanish. The same conclusion follows from $C_{2z}$.

By contrast, under $C_{2y}$, the longitudinal response remains compatible with the symmetry operation. Therefore, in pristine point group $T$, a nonzero longitudinal odd-in-$B$ response in Voigt geometry is forbidden specifically because the two $C_2$ axes perpendicular to $\mathbf{B}$ enforce cancellation. A nonzero response requires breaking both orthogonal $C_2$ symmetries while retaining the subgroup compatible with the field direction.

\subsection{Defect ensemble at $B=0$}
\label{SI.NoteS1.C}

We now consider a defect ensemble in which each local defect induces a local polar vector \(\vb F\) and possibly a local magnetic moment, while the set of symmetry-related defect sites remains statistically uniform. Denoting the ensemble average by \(\avg{\cdots}\), the odd moments of \(\vb F\) vanish,
\begin{equation}
\avg{F_i}=0,\qquad
\avg{F_iF_jF_k}=0,
\end{equation}
while the even moments reduce to isotropic tensor contractions,
\begin{equation}
\avg{F_iF_j}=\frac{F^2}{3}\delta_{ij},\qquad
\avg{F_iF_jF_kF_\ell}=A(\delta_{ij}\delta_{k\ell}+\delta_{ik}\delta_{j\ell}+\delta_{i\ell}\delta_{jk}),
\end{equation}
with $A$ fixed by \(\avg{F^4}\).

For a generic current expansion from Eq.~\ref{MPVE},
\begin{equation}
J_i(\vb E,\vb B;\vb F)
=\sum_{j,l\ge 0}
\chi^{(2,j;l)}_{\,i;\,(\alpha_1\alpha_2);(\beta_1\cdots\beta_j);(\gamma_1\cdots\gamma_l)}
E_{\alpha_1}E_{\alpha_2}
\left(\prod_{a=1}^{j} B_{\beta_a}\right)\left(\prod_{b=1}^{l} F_{\gamma_b}\right),
\qquad i,\alpha_r,\beta_j,\gamma_l\in\{x,y,z\},
\label{eq:J-with-F}
\end{equation}
all contributions with odd powers of \(\vb F\) vanish after ensemble averaging, while even powers reduce to isotropic renormalizations of already allowed terms. Therefore, at $B=0$, defect averaging cannot by itself activate a response channel that is symmetry-forbidden in the pristine $T$ crystal. It can only renormalize the strength of allowed responses.

\refstepcounter{section}
\section*{Note \thesection. How spins order at $B=0$: determining the easy axis of the charged oxygen vacancy}
\label{SI.NoteS2}

The charged oxygen vacancy hosts a local moment whose orientation is set by MAE induced by SOC.
We determine the MAE landscape by constraining the magnetization direction $\hat{\vb m}$ (unit vector) at $B=0$ and computing the total energy $E^{(i)}(\hat{\vb m})$ for each defect configuration $i$.

To sample orientations uniformly on the sphere, we use a spiral (Fibonacci-type) lattice on $\mathbb{S}^2$.
We generate $N_h$ points on the upper hemisphere ($z\ge 0$) and include their antipodal partners to enforce the reversal symmetry
\begin{equation}
E(\hat{\vb m})=E(-\hat{\vb m}),
\label{eq:even_property}
\end{equation}
expected at $B=0$.
With $i=0,1,\dots,N_h-1$,
\begin{equation}
z_i = \frac{i+\tfrac{1}{2}}{N_h},\qquad
\theta_i = \arccos(z_i),\qquad
\phi_i = 2\pi\,\mathrm{frac}\!\left(\frac{i}{\varphi_g}\right),
\label{eq:fib_angles}
\end{equation}
where $\varphi_g=(1+\sqrt{5})/2$ is the golden ratio.
The corresponding unit vectors are
\begin{equation}
\hat{\vb m}_i =
(\sin\theta_i\cos\phi_i,\ \sin\theta_i\sin\phi_i,\ \cos\theta_i),
\label{eq:fib_vectors}
\end{equation}
and the full antipodal set is
\begin{equation}
\mathcal{S}
=
\left\{ \hat{\vb m}_i \right\}_{i=0}^{N_h-1}
\cup
\left\{ -\hat{\vb m}_i \right\}_{i=0}^{N_h-1},
\label{eq:antipodal_set}
\end{equation}
containing $N=2N_h$ directions.
In this work we use $N_h=27$ (hence $N=54$ total directions).

\subsection*{Quadratic (leading-anisotropy) fit and symmetry covariance}
We first fit the MAE using the quadratic form
\begin{equation}
  E^{(i)}(\hat{\vb m}) = C^{(i)} + \hat{\vb m}^{\mathsf T} M^{(i)} \hat{\vb m},
  \qquad \hat{\vb m} = \frac{\vb m}{\|\vb m\|},
  \label{eq:MAE_quad}
\end{equation}
where $M^{(i)}$ is a real symmetric $3\times 3$ matrix.
Because $\hat{\vb m}^{\mathsf T}(\alpha I)\hat{\vb m}=\alpha$ for unit vectors, $M\to M+\alpha I$ can be absorbed into $C$.
The fitted easy axis $\hat{\vb m}^{(i)}_{\mathrm{easy}}$ is obtained from the eigenvector associated with the minimum eigenvalue of $M^{(i)}$.

Across the twelve symmetry-related vacancy configurations, the quadratic-fit easy axes transform covariantly under the crystallographic rotations: if configuration 2 is obtained from configuration 1 by a spatial rotation $R$, then
\begin{equation}
\hat{\vb m}^{(2)}_{\mathrm{easy}} \approx R\,\hat{\vb m}^{(1)}_{\mathrm{easy}}.
\label{eq:easy_axis_covariance_quad}
\end{equation}
Table~\ref{m_easy} reports representative examples; the full list is provided separately.

\begin{table}[h]
  \centering
  \caption{Representative quadratic-fit easy axes $\hat{\vb m}_{\mathrm{easy}}$ for charged oxygen-vacancy configurations. Components are in the global Cartesian frame.}
  \label{m_easy}
  \begin{tabular}{c|rrr}
    \toprule
    tag & $m_{\mathrm{easy},x}$ & $m_{\mathrm{easy},y}$ & $m_{\mathrm{easy},z}$  \\
    \midrule
    01 & 0.733 & 0.141 & 0.665 \\
    02 & 0.794 & -0.282 & -0.539 \\
    03 & -0.744 & 0.158 & -0.646 \\
    04 & -0.702 & -0.113 & 0.702 \\
    05 & 0.185 & 0.641 & 0.745 \\
    06 & 0.154 & -0.703 & -0.693 \\
    07 & -0.156 & 0.652 & 0.741 \\
    08 & -0.134 & -0.705 & 0.696 \\
    09 & 0.667 & 0.731 & 0.137 \\
    10 & 0.616 & -0.773 & -0.147 \\
    11 & -0.697 & 0.709 & -0.099 \\
    12 & -0.705 & -0.687 & 0.173 \\
    \bottomrule
  \end{tabular}
\end{table}

\refstepcounter{section}
\section*{Note \thesection. Geometric content of the TR-breaking photocurrent channels}
\label{SI.NoteS3}

In this section we make explicit that the two dominant time-reversal-breaking (TRb) contributions in the main text,
namely the circular shift current ($\sigma_{\mathrm{TRb}}$) and the linear injection current
($\eta_{\mathrm{TRb}}$), are correlated with Berry-curvature-like and quantum-metric-like band-geometric objects,
respectively. The starting point is the decomposition
Eqs.~(\ref{SC_decom}) and (\ref{IC_decom}) in main text, reproduced here for convenience. 
For complex interband connections $r^b_{nm}(\mathbf{k})$ and $r^c_{mn}(\mathbf{k})$, define the product
\begin{equation}
P^{bc}_{mn}(\mathbf{k}) \equiv r^b_{nm}(\mathbf{k})\,r^c_{mn}(\mathbf{k}).
\end{equation}
The commutator and anticommutator satisfy the identities
\begin{align}
[r^b_{nm},r^c_{mn}]
&= r^b_{nm}r^c_{mn}-r^c_{nm}r^b_{mn}
= P^{bc}_{mn} - \big(P^{bc}_{mn}\big)^{*}
= 2i\,\mathrm{Im}\,P^{bc}_{mn}, \label{SI:comm_im}\\
\{r^b_{nm},r^c_{mn}\}
&= r^b_{nm}r^c_{mn}+r^c_{nm}r^b_{mn}
= P^{bc}_{mn} + \big(P^{bc}_{mn}\big)^{*}
= 2\,\mathrm{Re}\,P^{bc}_{mn}. \label{SI:anticomm_re}
\end{align}
Therefore, the injection-current decomposition in Eq.~(\ref{IC_decom}) cleanly separates the antisymmetric
(helicity-odd) and symmetric (helicity-even) interband-coherence structures via
$[r^b_{nm},r^c_{mn}]$ and $\{r^b_{nm},r^c_{mn}\}$.

\subsection{Linear injection current is governed by the band quantum metric}
\label{SI:inj_metric}

The quantum metric of band $n$ is defined as the symmetric part of the quantum geometric tensor,
\begin{equation}
g^{(n)}_{bc}(\mathbf{k})
\equiv
\mathrm{Re}
\sum_{m\neq n}
r^b_{nm}(\mathbf{k})\,r^c_{mn}(\mathbf{k}).
\label{SI:qm_def}
\end{equation}
Using Eq.~(\ref{SI:anticomm_re}), the TR-breaking injection coefficient in Eq.~(\ref{IC_decom}) becomes
\begin{align}
\eta^{abc}_{\mathrm{TRb}}
&=
-\,\frac{\pi e^{3}}{2\hbar^{2}}
\!\int\![d\mathbf{k}]\,
\sum_{mn\sigma}
f_{mn}\,\Delta^{a}_{mn}\,
\{r^b_{nm},r^c_{mn}\}\,
\delta(\omega_{nm}-\omega) \nonumber\\
&=
-\,\frac{\pi e^{3}}{\hbar^{2}}
\!\int\![d\mathbf{k}]\,
\sum_{mn\sigma}
f_{mn}\,\Delta^{a}_{mn}\,
\mathrm{Re}\!\big(r^{b}_{nm}r^{c}_{mn}\big)\,
\delta(\omega_{nm}-\omega).
\label{SI:etaTR_metric_like}
\end{align}
Equation~(\ref{SI:etaTR_metric_like}) shows that the linear-in-polarization injection channel
($\propto \mathrm{Re}(E_bE_c^{*})$ in Eq.~(\ref{IC_decom})) is governed by the symmetric interband coherence
$\mathrm{Re}(r^b_{nm}r^c_{mn})$, i.e., the same geometric structure that builds the quantum metric
Eq.~(\ref{SI:qm_def}). In this sense, $\eta^{abc}_{\mathrm{TRb}}(\omega)$ defines an optical quantum metric
weighted by the velocity difference $\Delta^a_{mn}$ and the resonance constraint $\delta(\omega_{nm}-\omega)$.

\subsection{Circular shift current is governed by the band Berry curvature}
\label{SI:shift_bc}

The interband Berry curvature associated with the $(m,n)$ subspace can be written as
\begin{equation}
\Omega^{d}_{mn}(\mathbf{k})
=
i\,\epsilon^{dbc}\,r^{b}_{nm}(\mathbf{k})\,r^{c}_{mn}(\mathbf{k}),
\label{SI:bc_def}
\end{equation}
which is explicitly antisymmetric in the optical indices $(b,c)$ and originates from the imaginary part of the
interband coherence. Substituting Eq.~(\ref{SI:bc_def}) into the TR-breaking shift coefficient in Eq.~(\ref{SC_decom}),
we obtain
\begin{align}
\sigma^{abc}_{\mathrm{TRb}}
&=
-\,\frac{\pi e^{3}}{4i\,\hbar^{2}}
\!\int\![d\mathbf{k}]\,
\sum_{mn\sigma}
f_{nm}\,
\big(R^{a;b}_{mn}+R^{a;c}_{mn}\big)\,
\Omega^{d}_{mn}\,
\delta(\omega_{nm}-\omega) \nonumber\\
&=
-\,\frac{\pi e^{3}}{4\hbar^{2}}
\!\int\![d\mathbf{k}]\,
\sum_{mn\sigma}
f_{nm}\,
\big(R^{a;b}_{mn}+R^{a;c}_{mn}\big)\,
\epsilon^{dbc}\,r^{b}_{nm}r^{c}_{mn}\,
\delta(\omega_{nm}-\omega).
\label{SI:sigmaTR_bc_like}
\end{align}
Equation~(\ref{SI:sigmaTR_bc_like}) makes explicit that the circular shift-current channel
($\propto \mathrm{Im}(E_bE_c^{*})$ in Eq.~(\ref{SC_decom})) is governed by the antisymmetric interband coherence
encoded in $\Omega^{d}_{mn}(\mathbf{k})$, i.e., the Berry-curvature-like structure. Here the shift-vector prefactor
$\big(R^{a;b}_{mn}+R^{a;c}_{mn}\big)$ acts as a gauge-invariant weight selecting where the Berry-curvature hot spots
contribute most strongly at a given photon energy.

Combining Eqs.~(\ref{SI:etaTR_metric_like}) and (\ref{SI:sigmaTR_bc_like}), we conclude:
(i) the TR-breaking linear injection current is controlled by the symmetric (metric-like) interband coherence
$\mathrm{Re}(r^b_{nm}r^c_{mn})$ and thus correlates with the band quantum metric;
(ii) the TR-breaking circular shift current is controlled by the antisymmetric (curvature-like) interband coherence
$\Omega^{d}_{mn}$ and thus correlates with the band Berry curvature.
This provides the formal basis for comparing the k-resolved photocurrent hot spots with the k-resolved quantum metric
and Berry curvature maps in the main text.

\end{document}